%% file: main.tex
\documentclass[acmtog]{acmart}

\usepackage{algorithm}
\usepackage{algorithmic}

\usepackage{wrapfig}

\usepackage{multirow}
\usepackage{pgfplots}
\pgfplotsset{compat=1.9}

\setcopyright{none}
\renewcommand\footnotetextcopyrightpermission[1]{} 
\AtBeginDocument{%
  }

\copyrightyear{2023}
\acmYear{2023}

\acmMonth{4}

\input{preamble.tex}

\begin{document}

\title{Synchronized-tracing of implicit surfaces}

\author{C\'{e}dric Zanni}
\email{cedric.zanni@loria.fr}
\affiliation{%
  \institution{Universit\'{e} de Lorraine, CNRS, Inria, LORIA}
  \city{Nancy}
  \country{France}
}


\begin{abstract}
    Implicit surfaces are known for their ability to represent smooth objects of arbitrary topology thanks to hierarchical combinations of primitives using a structure called a blobtree. We present a new tile-based rendering pipeline well suited for modeling scenarios, i.e., no preprocessing is required when primitive parameters are updated.
    When using approximate signed distance fields, we rely on compact, smooth CSG operators - extended from standard bounded operators - to compute a tight volume of interest for all primitives of the blobtree.
    The pipeline relies on a low-resolution A-buffer storing the primitives of interest of a given screen tile. The A-buffer is then used during ray processing to synchronize threads within a subfrustum. This allows coherent field evaluation within workgroups. We use a sparse bottom-up tree traversal to prune the blobtree on-the-fly which allows us to decorrelate field evaluation complexity from the full blobtree size. The ray processing itself is done using the sphere-tracing algorithm.
    The pipeline scales well to surfaces consisting of thousands of primitives.
\end{abstract}

\ccsdesc[500]{Computing methodologies~Ray tracing}
\ccsdesc[500]{Computing methodologies~Volumetric models}

\keywords{implicit surfaces, blobtree, ray-tracing}

\begin{teaserfigure}
  \centering
  \includegraphics[width=0.95\textwidth]{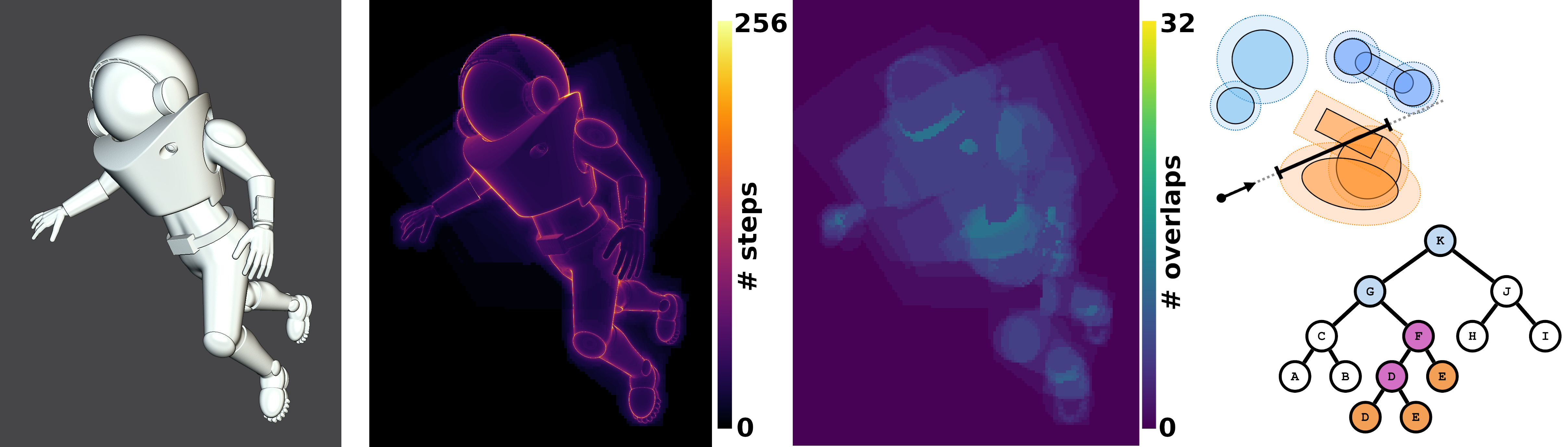}
  \caption{\label{fig:teaser}
  Approximate signed distance field (A-SDF) is a functional representation of volumes that allows modeling without concerns for topology. A-SDF can be modeled simply by the hierarchical composition of base primitives with blending and CSG operators (blobtree representation).
  We propose a new implicit rendering pipeline well suited for modeling. The proposed pipeline scales well to trees consisting of thousands of primitives thanks to a sparse expression tree traversal and synchronized ray processing.
  \emph{Left:} Astronaut model consisting of 142 primitives combined with smooth unions, intersections and differences.
  \emph{Middle left:} Number of field evaluations per ray.
  \emph{Middle right:} Maximal number of primitives overlap processed per screen tile. 
  \emph{Right:} When processing a ray subrange, we want the field evaluation complexity to be proportional to the number of primitive overlaps and not the full blobtree size.
}   
\end{teaserfigure}

\maketitle


\input{introduction.tex}

\input{previous_work.tex}

\input{overview.tex}

\input{core_blobtree.tex}

\input{core_blend.tex}

\input{core_acceleration.tex}

\input{core_sparse_traversal.tex}

\input{core_synchronized.tex}

\input{core_interdiff.tex}

\input{implementation.tex}

\input{discussion.tex}

\input{appendix_comparison.tex}

\input{conclusion.tex}

\begin{acks}
  This work was supported by the ANR IMPRIMA(ANR-18-CE46-0004).
  We thank Sylvain Lefebvre, Xavier Chermain and David Jourdan for the helpful discussion during the writing process.
\end{acks}

\bibliographystyle{ACM-Reference-Format}
\bibliography{implicit}


\end{document}

%% file: preamble.tex
\newcommand{\vr}{\mathbf{r}}
\newcommand{\vp}{\mathbf{p}}

\DeclareMathOperator\sign{sign}

\newcommand\R{\mathbb{R}}



%% file: introduction.tex
\section{Introduction}
Implicit surfaces - or volumes - are defined as set of point $\vp \in \R^3$ verifying the equation $f(\vp)\leq c$ (or $f(\vp)\geq c$ depending on the convention used) where $f$ is a user-defined scalar function (usually continuous and smooth by part) and $c$ a fixed iso-value. They are known for their ability to represent smooth objects of arbitrary topology.

 Among them, a large family of representations is built on the combination of simple base primitive functions, such as implicit volume representing spheres and cubes, combined through operators such as sharp union and smooth blending. Operations can be chained to define a construction tree, usually called a Blobtree~\cite{Wyvill97b,Wyvill99}. The tree leaves are primitives, and non-leaf nodes are operators. This representation can cover main implicit surface conventions: density fields~\cite{Wyvill86} and functional representations~\cite{Pasko95}, a superset of signed distance fields. The latters are used in an extensive range of applications, including CAD packages~\cite{ntopology,libfive}, demoscene~\cite{burger2002realtime,shadertoy}, modeling applications~\cite{magicacsg} and games~\cite{clay,dreams}.

Implicit surfaces are typically expensive to visualize and therefore are generally meshed or voxelized in modeling frameworks.  
If interactive feedback is required, a low resolution has to be used, which leads to a loss of details and poor edge reconstruction.
Ray-tracing algorithms, including the well-known sphere-tracing algorithm~\cite{Hart1996}, can alleviate these precision problems. However, when used on GPU, the expression of the implicit surface is often directly encoded inside the ray-tracing shader. As a result, large expressions that can arise in complex scenes can drastically increase rendering time and shader compilation time, hindering usage during modeling stages where the expression evolves frequently.

Recently, a large body of work took an interest in direct visualization algorithms that are data-oriented, i.e., that do not always require shader recompilation when the object is modified.
Most of those works are targeted at a single type of primitives combined with a unique n-ary summation blending operator~\cite{Bruckner2019,Aydinlilar2021}. The commutative nature of the summation blend allows for sparse expression evaluation from a subset of primitives present in the scene; this allows decorrelating the field evaluation's complexity from the full expression size. Similar property is desirable for blobtree evaluation (see Figure~\ref{fig:teaser}).
Notable exceptions are the approach proposed by~\citet{grasberger2016,Keeter2020} to process efficiently more general implicit surfaces on the GPU. 
Those methods represent the whole expression as data (a tree or tape recording the operations) that will be interpreted on the fly on the GPU without requiring shader recompilation when the model (i.e., the data) changes.
However, those two approaches rely either on progressive expression simplification (on the fly during evaluation) or static acceleration structures (as preprocessing). Therefore, both strategies require expensive processing of the entire tree to limit the complexity of subsequent field evaluation, hindering usage in modeling contexts.

\paragraph*{Our contributions} 
In this work, we take inspiration from all aforementioned data-oriented approaches. Our contributions focus on data management and field evaluation for efficient GPU processing. We propose a new implicit rendering pipeline that decorrelates field evaluation complexity from full blobtree size while avoiding preprocessing the full tree during base modeling operations (i.e., primitive updates). We focus on approximate signed distance fields (A-SDF), but the overall GPU pipeline could be used for density fields without significant changes. We take inspiration from compactly supported density fields to design compact, smooth constructive solid geometry (CSG) operators for A-SDF, which allow simple computation of tight volume of interest for primitives. We use those volumes to build a tile-based view-dependent acceleration structure. 
This structure is then used for workgroup synchronization and coherent access to primitives of interest. Based on the primitive of interest, we propose to prune the tree on the fly using a sparse bottom-up tree traversal only accessing a subset of blobtree nodes, those decorrelating the field complexity from the full expression size (see Figure~\ref{fig:teaser}).

We demonstrate the efficiency of our approach using a sphere-tracing algorithm on various examples, including large blobtrees and procedural nodes. The proposed pipeline scales well to trees consisting of thousands of primitives.


%% file: previous_work.tex
\section{Related work}

\subsection{Implicit Modeling}

Implicit surface modeling is a long-studied paradigm taking its root in constructive geometry
with the functional definition of smooth and sharp union and intersection~\cite{Ricci73}.
An in-depth introduction to the paradigm is available in~\cite{Bloomenthal97}.

Two main conventions are used for the field definition.
On the one hand, \emph{density fields} are similar to a density of materials that fall off to zero when going away from the shape, used with $c>0$. Density fields can naturally encompass radius of influence to localize shape influence in space (i.e. compactly supported primitives). This family provides the most straightforward smooth blending operator, i.e., the summation~\cite{Wyvill86}. 
On the other hand, \emph{functional representation}~\cite{Pasko95} has field values that increase toward infinity when going away from the shape and are used with $c=0$.
A notable subset of this family are signed distance fields (SDF), and approximate ones (A-SDF), where the field is defined as the (approximate) signed distance to the shape. SDFs are interesting for processing thanks to a Lipschitz bound equal to $1$. The Lipschitz bound is an upper bound on the norm of the field gradient.

In both cases, complex shapes can be defined in a constructive way by combining simpler primitives through a hierarchical blend tree called a blobtree~\cite{Wyvill99} for density fields. We will use this naming convention throughout the text.
For density field and A-SDF, sharp CSG union and intersection can be defined thanks to minimum and maximum functions. For example, for A-SDF, the union between two implicit volumes - defined by field $f_0$ and $f_1$ - can be defined as:
\begin{equation}
    csg_\cup(f_0,f_1) = \min(f_0, f_1) \ ,
    \label{eq:union}
\end{equation}
the intersection as:
\begin{equation}
    csg_\cap(f_0,f_1) = \max(f_0, f_1)  \ ,
    \label{eq:intersect}
\end{equation}
and the difference as:
\begin{equation}
    csg_\setminus(f_0,f_1) = \max(f_0, -f_1) \  .
    \label{eq:diff}
\end{equation}
Note that alternative formulations allow to define smooth field outside the iso-surface of interest.

\begin{figure}[b]
    \centering
    \includegraphics[width=0.45\linewidth]{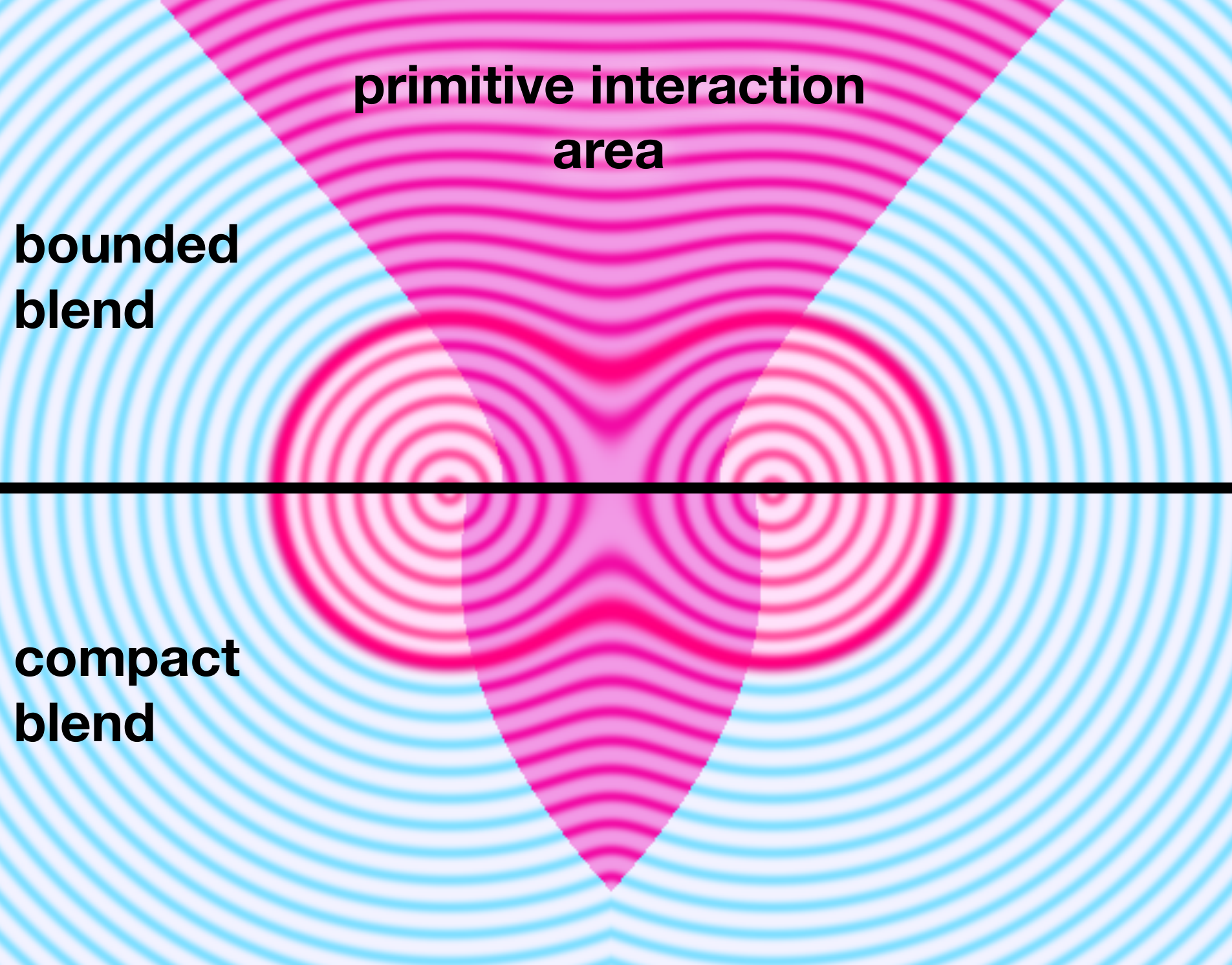}
    \caption{\label{fig:opertor-with-primitive}  \emph{Up:} Standard bounded smooth blending operators modify field value in an infinite volume (i.e., they are bounded at the 0 iso-value level only). This either increases field evaluation complexity or requires additional preprocessing. \emph{Down:} Inspired by density fields, we propose to localize fully blending regions in space. }
\end{figure}

An extensive range of operators has been introduced to create smooth, controllable blends between shapes for both conventions.
Many blending operators targeting functional representation provide broadly adopted bounded blending property~\cite{Pasko2002,Pasko2005,dekkers2004combining}.
A bounded blend localizes the blend in a bounded volume around the intersection of the 0 iso-surface of the two shapes to combine. However, the interaction between the two blended shapes is not bounded in space, i.e., both field operands can be required to compute the blended field far from the unblended shapes (see Figure~\ref{fig:opertor-with-primitive}). Bounded blends are usually defined as a displacement applied to a CSG operator: 
$$g(f_0,f_1) = CSG(f_0,f_1) \pm disp(f_0,f_1) \ . $$
The lack of field interaction localization is either due to the usage of clean CSG R-function operators \cite{Pasko2002,Pasko2005} to produce smooth field function everywhere in space or to an unbounded displacement function to prevent an increase of the Lipschitz bound of the field~\cite{dekkers2004combining}.

In the present work, to simplify the processing of large blobtrees, we propose strengthening the bounded blend property to guarantee bounded interaction volume similar to the one provided by compactly supported density fields.
We will base our modified operators on a family of smooth polynomial bounded operator~\cite{dekkers2004combining} providing a simple closed-form definition well suited for massive GPU evaluation that has been used for games~\cite{dreams} and demoscene~\cite{iquilezles}. The $C^2$ variant for the smooth union is defined as:
\begin{equation}
    g_\cup(f_0,f_1,k) = csg_\cup(f_0, f_1) - d(f_0,f_1,k) 
    \label{eq:smooth-union}
\end{equation}
with 
\begin{equation}
    d(f_0,f_1,k) = \frac{k}{6} \max \left( 1-\frac{|f_0-f_1|}{k}, 0\right)^3
    \label{eq:disp_smooth}
\end{equation}
where $k$ is a parameter defining the blend range.
The smooth intersection is similarly defined as:
\begin{equation}
    g_\cap(f_0,f_1,k) = csg_\cap(f_0, f_1) + d(f_0,f_1,k) 
    \label{eq:smooth-intersect}
\end{equation}
and the smooth difference as:
\begin{equation}
    g_\setminus(f_0,f_1,k) = csg_\setminus(f_0, f_1) + d(f_0,-f_1,k)
\label{eq:smooth-diff}
\end{equation}

\input{Z_pipeline.tex}

\subsection{Implicit volume visualization}

Direct visualization of implicit surfaces requires solving ray iso-surface intersection for each pixel of an image:
$$f \circ \vr(t)=c$$
where $\vr(t)$ is the ray parameterization of a given pixel ($c=0$ for A-SDF). 
Doing so efficiently requires studying three aspects: the root-finding problem itself, the representation of the field function, and acceleration strategies when a direct field evaluation is expensive.

\paragraph{Ray processing routines}
The simplest but expensive root finding strategy is to ray march along the ray with a constant step size until a sign change is detected; this leads to a tradeoff between significant rendering time and rendering artifacts (missed details).

In order to limit the number of field evaluations, a first family of methods relies on polynomial field representation~\cite{Nishita1994,Gourmel-2010} or approximation~\cite{Sherstyuk1999RT,Aydinlilar2021} restricting the range of application to a specific type of implicit surfaces. 

A more general class of method relies on ray bisection driven by inclusion function bounding $f$ on a ray interval. Inclusion functions are build using either interval arithmeric~\cite{Mitchell1990,duff1992interval,knoll2009fast,Keeter2020} or variant of affine arithmetic~\cite{knoll2009fast,Fryazinov2010RI,sharp2022spelunking}. 
The stack required to manage the recursion can be a limiting factor for GPU implementation.
A notable exception is~\cite{Keeter2020} which approach is more akin to screen space voxelization.

The last class of method relies on the Lipschitz bound of $f$ to compute intersection free step-size and advance iteratively along the ray. The seminal sphere-tracing algorithm~\cite{Hart1996} uses either a global bound or localized one relying on an octree.
 While overrelaxation can be used to reduce the number of steps~\cite{Keinert2014,Balint2018}, grazing rays still require many steps. 
 Segment-tracing~\cite{Galin2020} provides a solution by computing directional Lipschitz bound on ray subintervals. This method was presented for density fields, but nothing prevents its application to A-SDF. In \cite{seyb19}, a variant is dedicated to deformed surfaces.

In this paper, we rely on sphere-tracing for its simplicity, but we design the pipeline architecture for iterative processing methods in general.

\paragraph{Representation of field expression} 
When implementing a ray processing routine on GPU, the encoding of the field expression has important implications. If represented as function calls directly hard coded inside a shader~\cite{reiner2011interactive,shadertoy}, good performance can be achieved for reasonable size expressions at the expense of shader recompilation each time the expression is modified.
Alternatively, the expression can be represented as data in GPU memory. When a unique primitive type and commutative operator are used~\cite{Gourmel-2010,Bruckner2019,Aydinlilar2021}, the entire expression can be stored as an array of primitives parameters.
For more general implicit surfaces - such as those represented by blobtrees - an interpreter can dispatch function calls based on a type identifier~\cite{grasberger2016,Keeter2020}. In order to avoid non-linear memory access and provide stackless expression/tree traversal, both methods rely on linearized data structures: post-order tree storage for blobtrees~\cite{grasberger2016} and tape for arithmetic expression represented as directed acyclic graph~\cite{Keeter2020}. Our data structure derives from~\cite{grasberger2016}, and we rely on a node-level interpreter to limit memory usage.

\paragraph{Acceleration strategies} 
For complex expressions, direct field evaluation can become prohibitively expensive. Several strategies have been studied to alleviate this problem. First, field values can be cached for re-use in subsequent processing~\cite{Schmidt2005,reiner2012}. This is of limited use if a shape undergoes significant modifications.

Most acceleration strategies rely on tree prunning~\cite{Fox2001} combined with spatial acceleration structures.
A pruned tree is a subtree that provides an equivalent expression for a given region in space. At field query time, the acceleration structure is used to access a pruned tree which can then be evaluated as usual. Those methods are best suited to density fields with compact supports, as pruned trees can drastically reduce the evaluation complexity. Both BVH~\cite{Gourmel-2010}, BSP-tree/kD-tree~\cite{grasberger2016}, and octree~\cite{dyllong2007verified} have been used for the acceleration structure. 
Such techniques require full tree preprocessing each time the primitive parameters (especially positions) are changed.
A BVH can also be used to stop top-down tree traversal on the fly~\cite{grasberger2016}, but this restricts the shape of the BVH tree; the latter and blobtree have opposed optimal tree shape (balanced versus left-heavy).

A completely different strategy relying on on-the-fly expression simplification is proposed by~\citet{Keeter2020}. The method uses a multigrid space subdivision to progressively simplify the expression in subcell using interval arithmetics. This also requires processing the whole expression in the first step of the rendering pipeline. A key feature of this approach is coherent processing, particularly well suited for GPU architecture, a property shared by our rendering pipeline.

When a single commutative operator is used, an A-buffer storing primitives of interest along each ray can be used~\cite{Bruckner2019,Aydinlilar2021}. We will adapt this solution to more general implicit surfaces represented by blobtrees, taking special care about ray processing coherency.

%% file: Z_pipeline.tex
\begin{figure*}[h]
    \centering
    \includegraphics[width=0.7\linewidth]{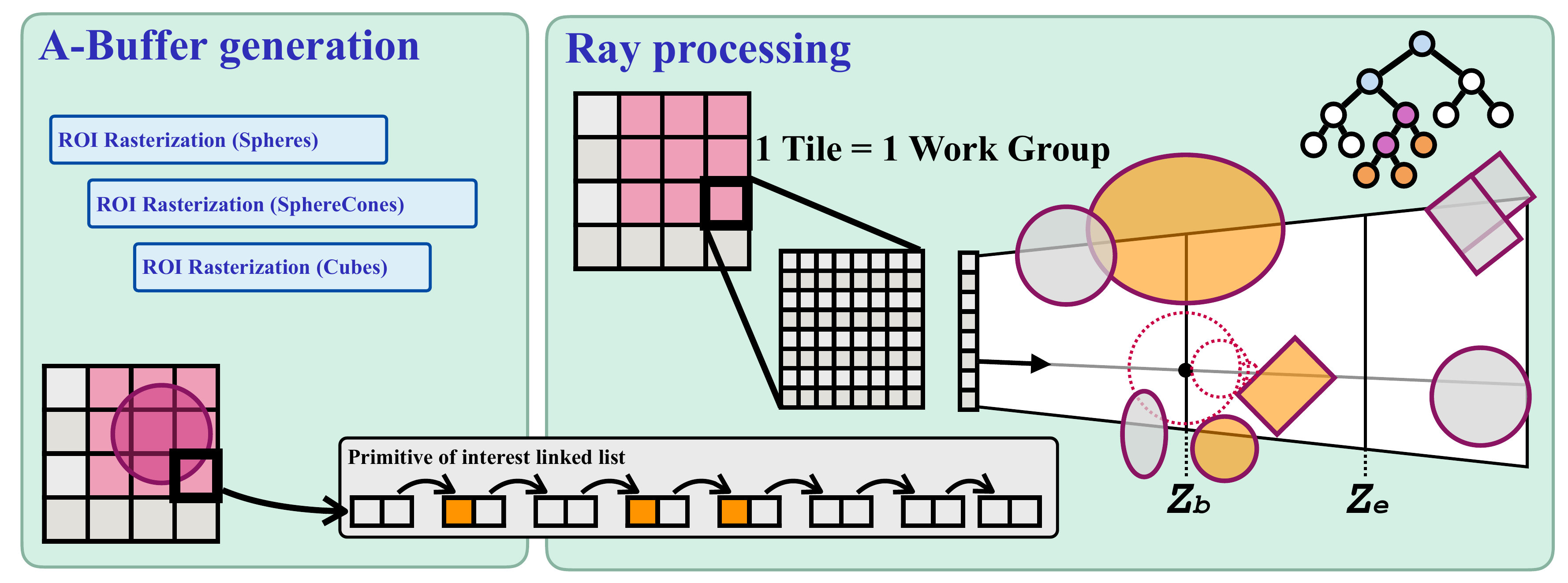}
    \caption{\label{fig:pipeline} Our rendering pipeline is tile-based and consist of two steps.
    \emph{left:} First, a rasterization step is used to populate linked-lists of a low resolution A-buffer with primitive volumes of interest (one list per screen tile).
    \emph{Right:} Then, a compute shader is used to process screen tiles (1 workgroup per tile). A linked-list is used to segment the rays of the corresponding screen tile. Threads of a workgroup therefore access to the blobtree in a coherent fashion. Using the active primitives of a subfrustum - colored in orange - we introduce a sparse blobtree evaluation to limit field evaluation complexity (see Figure~\ref{fig:teaser}). Ray subintervals are processed using the sphere tracing algorithm.
    }
\end{figure*}

%% file: overview.tex
\section{Overview}

We designed our rendering pipeline for coherent access to blobtree memory and field evaluation within GPU workgroups.
We, therefore, use a tile-based pipeline where rays are processed by buckets corresponding to screen tiles $T$ of size $8 \times 8$ pixels and an associated frustum $\mathcal{F}_T$.
Our pipeline consists of two main steps: first, the creation of a tile-based view-dependent structure storing a depth-sorted list of primitives of interest per tile (see Figure~\ref{fig:pipeline} - left); then, the actual ray processing to localize the first iso-surface crossing (see Figure~\ref{fig:pipeline} - right). 

\paragraph{Terminology} 
Our pipeline relies on the notion of local equivalence between A-SDFs in order to decrease field evaluation complexity. Given $f$ a reference A-SDF defined in $\R^3$, we define the local equivalence of an A-SDF $f_a$ with $f$ in the volume $\mathcal{V}$ as:
\begin{equation} 
        f_a \big|_\mathcal{V} \sim f \big|_\mathcal{V} \Leftrightarrow \forall \vp \in \mathcal{V} \ , \ \sign(f_a(\vp)) = \sign(f(\vp))
    \end{equation}
This implies that $f$ and $f_a$ define a common 0 iso-volume in $\mathcal{V}$. We define a primitive \emph{volume of interest} as the volume in which the primitive needs to be evaluated to compute an equivalent A-SDF from a subset of the blobtree nodes (i.e., a pruned blotree).     

Given a volume $\mathcal{V}$ to process, we define the set $\mathcal{A}$ of \emph{active primitives} as the set of primitives whose volume of interest intersects $\mathcal{V}$. Given the definition of volume of interest, a pruned version of the blobtree can then be used to define $f_\mathcal{A}$ such that $f_\mathcal{A}\big|_\mathcal{V} \sim f\big|_\mathcal{V}$. Contrary to tree pruning with compactly supported density fields, the pruned tree does not define the same field function; however, it is still an A-SDF so well suited for sphere-tracing. We propose a simple modification of standard smooth polynomial operators to obtain bounded volumes of interest when used with A-SDF primitives, which have infinite support.

\paragraph{Tiled acceleration structure} 
Given a screen tile $T_{ij}$, we define the associated \emph{primitives of interest} as the set of primitives whose volume of interest intersects the frustum $\mathcal{F}_{T_{ij}}$.
We use a tile-level A-buffer to track primitives of interest. We use a shared list of primitives per tile to simplify ray processing synchronization during the pipeline's second step. The low resolution of the A-buffer - downsampled by 8 in $x$ and $y$ - also provides several side benefits: lower memory usage and faster construction time.
The A-buffer is populated during the first step of the pipeline by low-resolution rasterization of the primitive volumes of interest. Registered information consists of an address in the blobtree and entry/exit depth inside a primitive volume of interest. Primitives are sorted by depth to simplify the next processing step.

\paragraph{Tiled ray processing} This step is performed in a compute shader; each workgroup corresponding to a screen tile $T_{ij}$. This allows us to synchronize ray traversal to achieve coherent access to blobtree data:
Primitives of interest registered in the A-Buffer are processed iteratively to segment the frustum $\mathcal{F}_{T_{ij}}$ into subfrustums $\mathcal{F}_{T_{ij}}[z_b,z_e]$ where a fixed set of \emph{active primitives} $\mathcal{A}$ is required to define $f_\mathcal{A}$. Finally, the sphere-tracing algorithm is applied for each subinterval until an iso-crossing is detected. 

The new field $f_\mathcal{A}$ is defined using a sparse bottom-up blobtree traversal only traversing branches terminated by active primitives (see Figure~\ref{fig:postorder_storage} \emph{top}). This traversal strategy can either be used for direct field evaluation or on-the-fly building of pruned blobtree views in GPU shared memory (as done in our complete pipeline).
\\

We first describe the representation used for the blobtree on GPU, and the associated node-level interpreters, in section~\ref{sec:blobtree_interpreter}.
We then discuss in section~\ref{sec:taxonomy_blend} an operator property allowing to localize trivially primitive influence in space. A simple extension of existing smooth union, intersection, and difference operators allows verifying this property.
The construction of the A-buffer is presented in section~\ref{sec:acceleration_structure}, and the sparse bottom-up tree traversal is presented in section~\ref{sec:bottom_up_eval}.
All aspects relating to synchronized processing are presented in section~\ref{sec:synchronized}.
For clarity, Sections~\ref{sec:taxonomy_blend} to~\ref{sec:synchronized} focus on the processing of union-like operators only. Section~\ref{sec:interdiff} then describes changes required to support intersection and difference-like operators properly.
Finally, results are presented in section~\ref{sec:discussion}.

%% file: core_blobtree.tex
\begin{figure}[bt]
    \centering
    \includegraphics[width=0.6\linewidth]{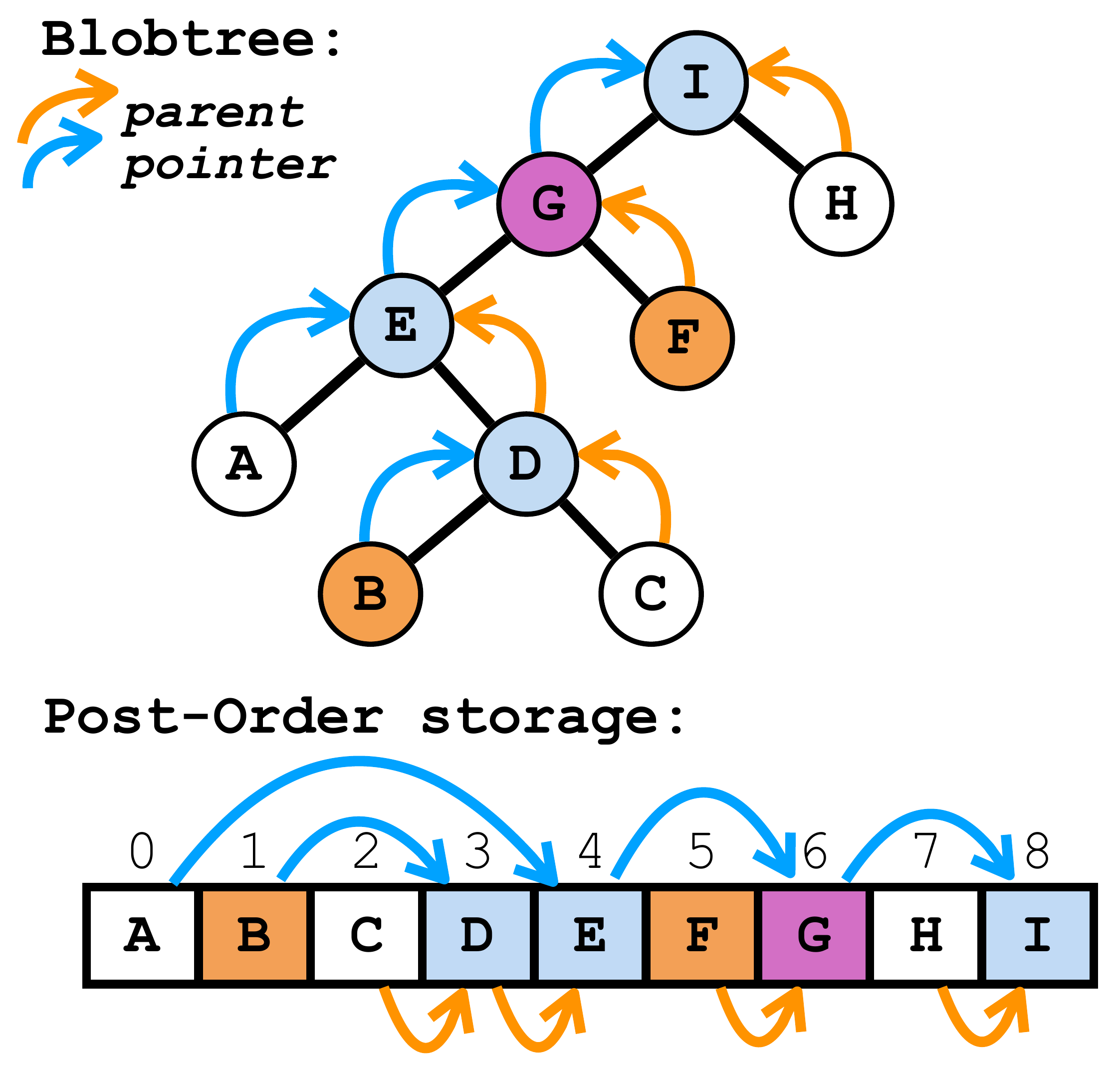}
    \caption{\label{fig:postorder_storage} \emph{Top:} A sparse blobtree traversal should only
    access colored nodes starting from active primitives (in orange).
    Nodes in blue should only act as operand selectors while nodes in purple should be evaluated as usual. This color convention is reused in all figures.
    \emph{Bottom:} Similarly to~\cite{grasberger2016}, we store on GPU a linearized version of the blobtree where nodes appear in \emph{post-order} tree traversal order. For each node, 
we store an ancestor pointer (initially parent pointer for the sake of explanation).
}
\end{figure}

\section{Blobtree representation and interpreters}
\label{sec:blobtree_interpreter}

In order to prevent large number of shader recompilations during modeling, our blobtree representation derives from the data-oriented approach of~\cite{grasberger2016}; we rely on a linearized post-order tree for storage and node-level interpreters for field evaluation. As it is a core component in our method, we review it in detail, including small variations that will be required by our sparse tree traversal in order to manage intersection and difference-like operators.

\paragraph*{The Blobtree} Our input is a blobtree with space deformation moved to the leaves (our implementation only supports rotation and translation of primitives). As in~\cite{grasberger2016}, we linearize the tree storage on GPU using postfix storage order (see Figure~\ref{fig:postorder_storage}). 
This storage layout removes the need for a traversal stack when a complete bottom-up tree traversal is performed as the nodes are processed in storage order. We will instead rely on this ordering during our sparse tree traversal and provide an alternative way to limit the stack memory requirements (see Section~\ref{sec:synchronized}). 
Let us call $B$ the floating-point array used to store all blobtree data: we store the node information representing the tree structure interleaved with node parameters (reinterpret cast are used to access integers and bitfields when needed). Node allocation is done with a word size $w$ of 128 bits for memory alignment and a smaller ancestor/parent pointer memory footprint.
The first 32 bits of each node are reserved for packed generic node information representing the tree structure and the node type.
We will refer to this bitfield as \emph{blob} and use the following layout:

 {\small
\begin{tabular}{|c|c|c|c|c|}
 \hline
\emph{isPrimitive}(1) & \emph{nodeop}(5) & \emph{ignoreMod}(2) & \emph{isLeft}(1) & \emph{ancestor}(23)   \\ 
 \hline
\end{tabular}
}\\
where the number in parenthesis denote the number of bits used for each property.
This compact representation allows fetching in a single memory read during tree traversal. The remaining 96 bits of the first word is used for padding for memory alignment of operator and primitive parameters.

Given a pointer $p$ to the beginning of a blobtree node, the \emph{blob} bitfield $b$ is accessed with $B[4p]$. We can query the following information:
\begin{itemize}
\item $isPrimitive(b)$ : whether the node is a primitive (leaf node)
\item $type(b)$ : \emph{nodeop} identifier representing the type of the node
\item $ignoreMod(b)$ : two bits used to encode the behavior to follow when an operand is missing during sparse traversal (management of intersection and difference)
\item $isLeft(b)$ : whether the node is a left child \item $ancestor(b)$ : parent/ancestor pointer, parent information can, in turn be accessed with $B[4ancestor(b)]$
\end{itemize}

Up to 32 different primitive types can be encoded in \emph{nodeop}, and the same number of operators can be managed. Ignore mode bits are only useful for operators; they could be combined to primitive \emph{nodeop} to raise the number of useable primitive types to 128 (we did not put this optimization in place). 
Finally, the ancestor is the pointer of the parent node (or of another ancestor for sparser traversal as discussed in Section~\ref{sec:bottom_up_eval}). Using 23 bits for this value combined with a stride based on the word size allows to index blobtree of up to 123MB. Note that the distance between successive nodes is not constant and depends on the node type. All diagrams will only show the \emph{blob} bitfields for readability.
Note that putting aside the interleaved data layout, as done in \cite{grasberger2016}, would allow indexing much larger trees at the cost of an additional memory read (indirection) per node to process.

\paragraph*{Node level interpreters} As in~\cite{grasberger2016}, we use node-level interpreters. 
We also use a similar interpreter to access the memory footprint of a given node list of parameters whenever needed. 
Compared to mathematical expression interpreter~\cite{Keeter2020}, node-level interpreters decrease the number of switch executions and require less storage of data (encoded by the function called). Furthermore, they can support fully procedural nodes.
The two main downsides are that it results in a more complex shader which needs to be recompiled if the family of primitive/operator is changed (shader can be initialized with a larger set of classical primitives and operators to alleviate this problem).

%% file: core_blend.tex
\section{Taxonomy of implicit blend}
\label{sec:taxonomy_blend}

Our goal is to decorrelate the complexity of field evaluation from the blobtree size and limit empty space processing.

\paragraph{Range of interest} 
In order to limit the complexity of field evaluations, we need to be able to localize the influence of primitives in space in order to discard preemptively unneeded ones based on the field query location. 
For a given blobtree node $n$, let us call the node \emph{range of interest} (or ROI):
$$R_{n} = (-\infty,d_{n}]$$
the range in which the node should be evaluated to be able to define properly $f_\mathcal{A}$ in an arbitrary volume: if the inputs are in $R_n$, the output should either be inside parent ROI - and exact - or outside of parent ROI.
For primitives, such ranges can be converted into volumes of interest in order to localize processing in space (i.e., simple offsetting of the bounding volume of the primitive 0 iso-volume - see Section~\ref{sec:bounding_volume}).

In order to limit the amount of preprocessing, we would like to compute primitive ROIs 1) in a single tree traversal for efficiency and 2) independently from the space embedding of primitives such that no preprocessing is required when only primitives change. 

Classical bounded operators present difficulties in order to compute such ROIs. For instance, for two spheres combined with $g_\cup$, the area of interaction between the primitives extends to infinity (see Figure~\ref{fig:opertor-with-primitive}). Primitive ROIs might therefore grow infinitely depending on both subsequent blend operations and primitives positions in space breaking our requirements.

We propose to define a \emph{compactness} property for operators: an operator should have a range of interest $R_o$ - deriving only from operator blend parameters - defined as the largest field value at which its operands can interact.
After this threshold, the result of the operator should only be a selection of one of the operands - up to a sign inversion to manage difference-like operators -  effectively preventing primitives interactions outside of $R_o$, which in turn will allow deducing an equivalent field evaluation from active operands only.

Some definition of CSG operators meets this property (see Equation~(\ref{eq:union},\ref{eq:intersect},\ref{eq:diff}) and  Figure~\ref{fig:opertor-with-primitive}). 
We will therefore define our operators by part as:
\begin{equation}
	G_{op}(f_0,f_1) =  
	  \left\{ 
	  \begin{array}{ll}
		\bar{g}_{op}(f_0, f_1)
		  \ \hbox{ if }\ f_0 \in R_{o} \hbox{ and } f_1 \in R_{o}, \\
		  csg_{op}(f_0, f_1) \ \hbox{ otherwise}.
	 \end{array}
	 \right. 
\label{eq:smooth_comp}
\end{equation}
with $op \in \{\cup, \cap, \setminus \}$, falling back on CSG operator when at least one operand is outside of the operator range $R_o$. Local operator $\bar{g}_{op}$ is defined on $R_{o}^2$ and satisfies continuity constraints with $csg_{op}$ at the boundary of the domain resulting in a continuous operator on $\R^2$.

With this definition, we can compute primitive ROIs by propagating node ROIs down the tree from the root (the ROI of the root is $(-\infty,0]$). 
By default, the ROI to propagate is simply defined as $R_o \cup R_p$ with $R_p$ the ROI of the parent node (i.e., the context in which the operator is used). 


\begin{figure}[t]
    \centering
    \includegraphics[width=0.95\linewidth]{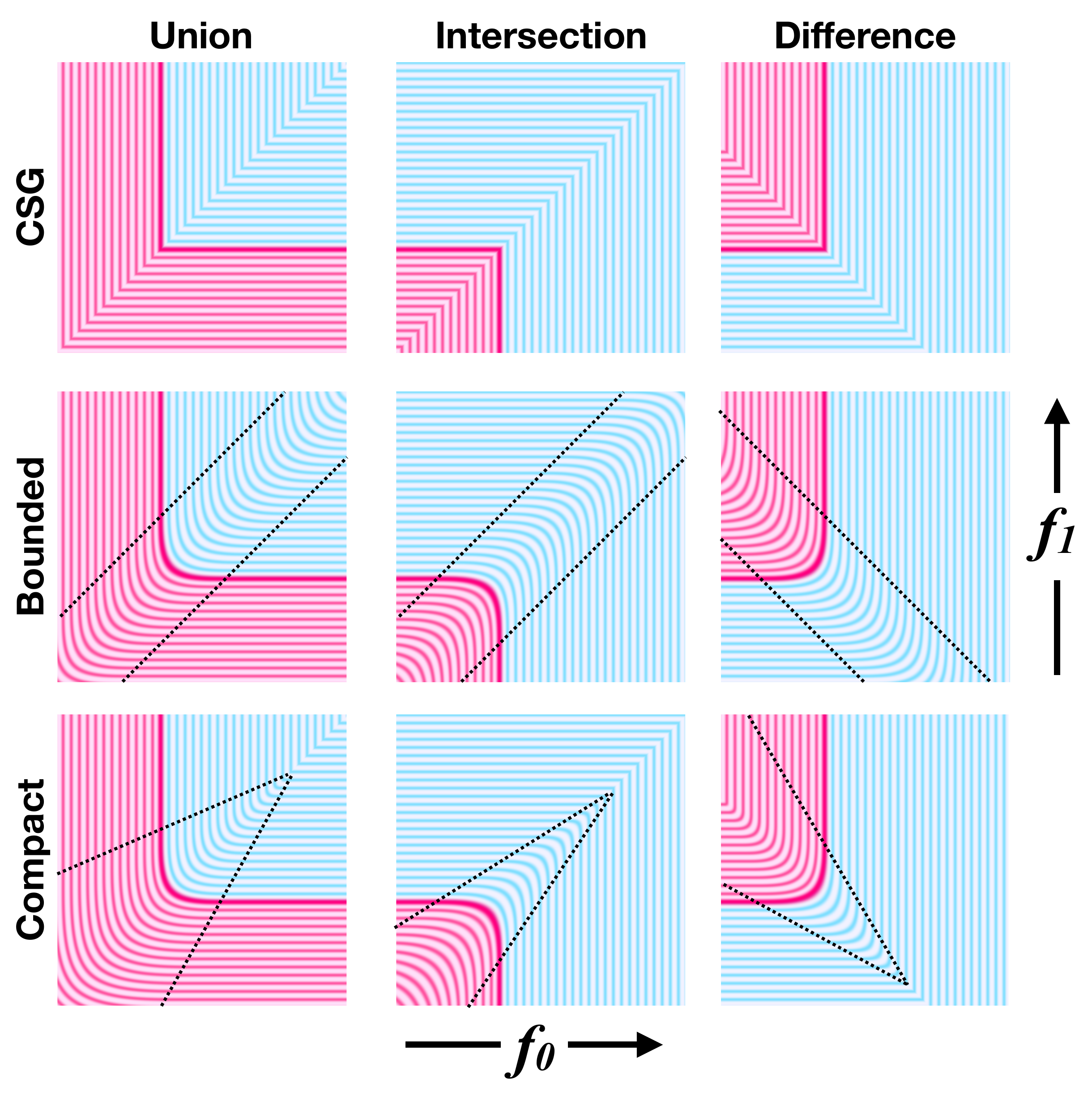}
    \caption{\label{fig:operator-compact} Blend graphs: operand $f_0$ and $f_1$ of an operator correspond to $x$ and $y$ axis. The interior of the resulting 0 iso-surface is depicted in red, exterior in blue. 
    \emph{Up:} CSG operators acting as operand selectors (return either one of the operands up to a minus sign).
    \emph{Middle:} Bounded smooth operators do not restrict operand interactions in space: both operands can be required to evaluate the operator at an arbitrary distance to the operands 0 iso-surface, 
    \emph{Down:} We propose a simple modification of smooth operators that localizes interaction between primitives in space. Outside of the interaction area, operators act as field selectors. 
}
\end{figure}


\paragraph{Evaluate operators outside of child ROIs:} Let consider what happens for a smooth union operator verifying Equation~(\ref{eq:smooth_comp}).
If both operands are active, the operator can be evaluated as usual. Let us assume we have $f_1 \notin R_{c_1}$ without loss of generality if at least one operand is outside of its range of interest. Then we have, $op(f_1,f_2,k) = min(f_1,f_2)$ and two cases can arise:
\begin{itemize}
\item $f_1 \leq f_2$, then we have $op(f_1,f_2) = f_1 > d_{c_1} \geq d_p$, which means that the parent will not require the result of the operator and an infinite value can be returned safely or an ignore status can be propagated in the tree.
\item $f_1>f_2$, then we have $op(f_1,f_2) = f_2$, if $f_2 \notin R_{c_2}$, the same reasoning can apply, else we can return $f_2$ which is the correct value for the operator.
\end{itemize} 
Similar reasoning can be followed for intersections and differences (see Section~\ref{sec:interdiff}).

Thanks to this property, given an arbitrary volume $\mathcal{V}$, an equivalent A-SDF field can be computed using only primitives whose volume of interest intersects $\mathcal{V}$. Hence, sphere-tracing can be used in $\mathcal{V}$.  


\paragraph{Adapting existing operators}
We only need to design operators $\bar{g}_{op}$ in the operator range $(-\infty, d_o]^2$ such that it continuously transition to CSG operators at the limit of the domain as described in Equation~(\ref{eq:smooth_comp}).
Standard polynomial smooth operators $g_\cup$, $g_\cap$, and $g_\setminus$ can be used as a base building block while preserving the same blend behavior when the operator is used in isolation.
We only have to progressively close the blend range when moving away from the $0$ iso-surface resulting from the unmodified blend such that the blend range is fully closed when reaching the limit of $(-\infty, d_o]^2$. 

This can be done by evaluating twice the unmodified operator, e.g., for the smooth union:
\begin{equation}
	\bar{g}_{\cup}(f_0,f_1) = g_\cup(f_0,f_1, b_r(g_\cup(f_0,f_1,k), k, d_o) )
	\label{eq:compact_smooth_union}	
\end{equation}
where the first evaluation is used to compute an interpolation of the blend parameter from $k$ on the 0 iso-surface of $g_\cup(f_0,f_1,k)$ to $0$ on the $d_o$ iso-surface (in order to switch to regular $\min$ union operator). Smooth intersection and difference use the same formula only replacing the smooth operator used by $g_\cap$ and $g_\setminus$, as well as using the absolute value of $b_r$ for the difference to properly limit the range of interaction.
For the transition function, we use the following:
$$b_r(x,k,d)=k \ \max\left( 1-\frac{6 x}{6 d-k}  , 0 \right)  $$
Note that a \emph{not a number} can appear during the evaluation of the blend range; it can be filtered by returning 0 instead.
The same formulation can be used for smooth intersection and difference; the only difference is that we set an upper bound equal to $k$ for the blend range resulting from the evaluation of $b_r$.
The resulting operators are depicted in Figure~\ref{fig:operator-compact}.

The blend operator, therefore, has an extra parameter $d_o$, which can either be set automatically based on $k$ or by hand. This can be useful to prevent significant field compression if a large number of primitive overlaps are expected (see discussion in Section~\ref{sec:discussion}).

%% file: core_acceleration.tex
\section{Acceleration structure}
\label{sec:acceleration_structure}

For a given subfrustum $\mathcal{F}_{T_{ij}}[z_b,z_e]$,
we need an efficient way to access active primitives (i.e., their pointers in $B$). Similarly, to~\cite{Bruckner2019,Aydinlilar2021}, which target density fields, we use an A-buffer built from primitive volumes of interest.

\subsection{Primitive volume of interest}
\label{sec:bounding_volume}

Assuming well-behaved primitives (Lipschitz bound smaller or equal to 1), we can build primitive volumes of interest by offsetting the transformed primitive bounding volume by the upper (positive) value of the range of interest. As explained in the previous section, ROIs are computed by propagation down the tree; 
this step is done on CPU.

To limit the number of primitive overlaps, the volumes of interest should be as tight as possible. We, therefore, do not restrict ourselves to a single type of volume but use several families of volumes (typically spheres, transformed cubes, and capsules). Each primitive type maps to a single volume type.

\subsection{A-buffer generation}

A-buffers are per-pixel - here per-tile - linked-list of fragments. Each fragment corresponds to primitive information - here, the primitive pointer inside $B$ - the entry depth inside the primitive volume of interest and the exit depth. Linked lists are sorted by entry depth.
The A-buffer is built by rasterization of the primitive volume of interest, using the parallel algorithm for insertion in sorted linked-list presented in~\cite{lefebvre:hal-01093158}. 

The main difference with~\cite{Bruckner2019,Aydinlilar2021} is the usage of an A-buffer, which entries correspond to screen tiles and not pixels. Using 8 by 8 tiles, drastically reduces the number of atomic operations during linked-list updates.

\paragraph*{Support rasterization:} We use the same approach for the rasterization of all types of volumes of interest. A rendering pass is performed at the downsampled resolution (number of screen tiles). We use a geometry shader to generate screen-space imposters, which rasterization will instantiate the computation of intersection between ray buckets and volume of interest. For each generated fragment, the fragment shader will compute $8^2$ ray/volume intersections and aggregate depth in a conservative way (min for entry, max for exit), then perform the insertion in the sorted list. For each volume of interest type, we store a mapping between the local index of a primitive within the family and the primitive pointer in $B$.

While we rely on three families of volumes (spheres, oriented-box, capsules), the system can be extended to a larger range of shapes such as cylinder, torus, sphere-cone, and sphere-triangle.

%% file: core_sparse_traversal.tex
\section{Sparse Bottom-up tree traversal}
\label{sec:bottom_up_eval}


\begin{figure}[tb]
    \centering
    \includegraphics[width=1.0\linewidth]{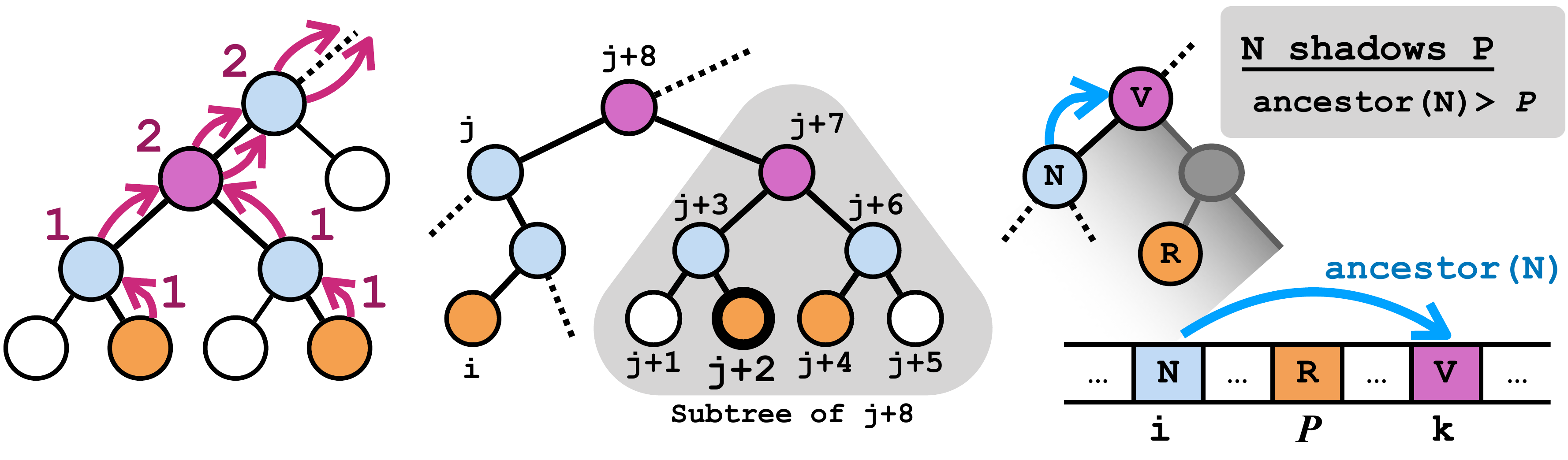}    

    \caption{\label{fig:shadowing}\emph{Left:} Traversing the tree from each active primitive to the root using parent pointers lead to multiple traversals of some operators (purple numbers correspond to number of traversal of a node).
    \emph{Middle:} The first node to process duringuipop sparse traversal of a subtree is always an active primitive.
    \emph{Right:} A node should not be fetched and processed if it is stored after the first unprocessed active primitive. We call this the shadowing condition (this condition is modified when using fast pointers).
    }
\end{figure}

\begin{figure}[tb]
    \centering
        \includegraphics[width=0.9\linewidth]{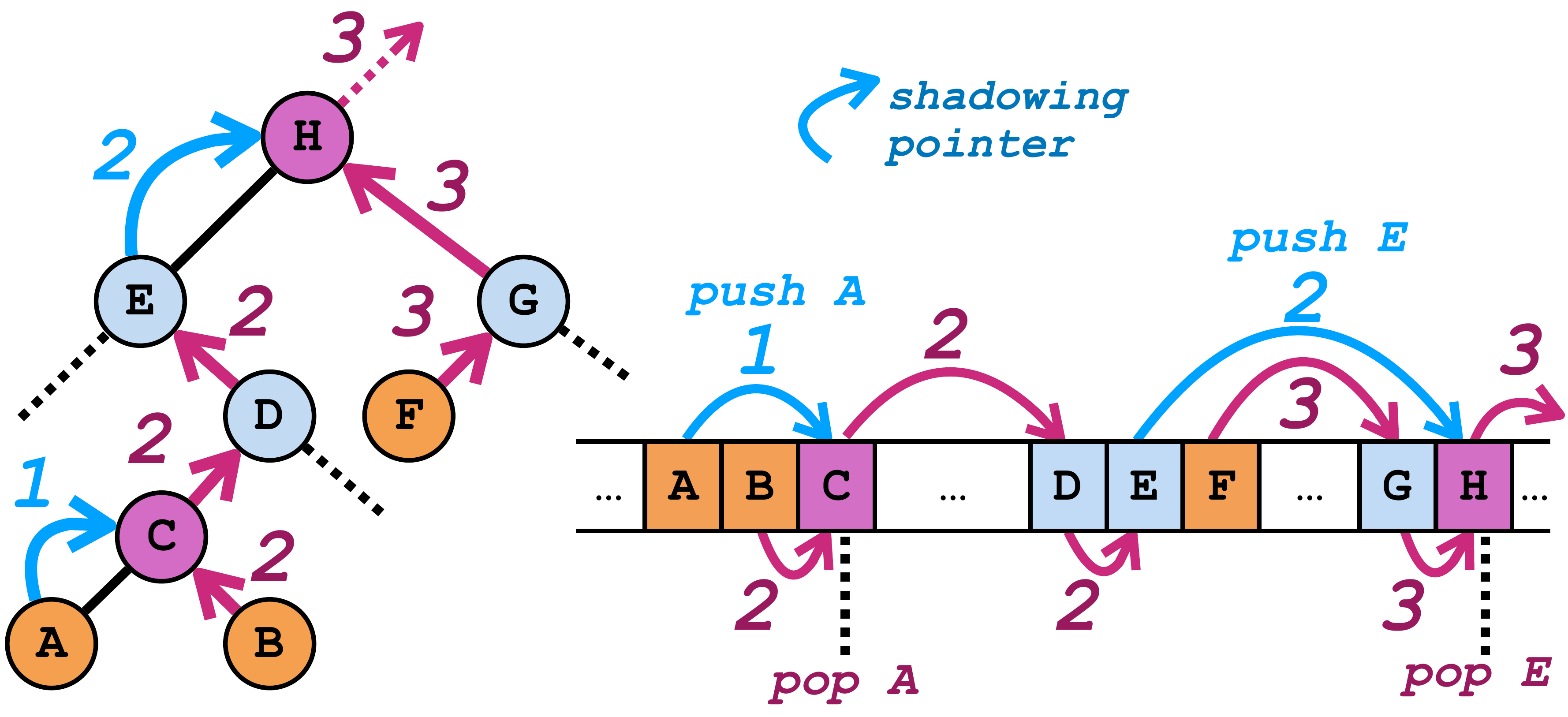}
    \caption{\label{fig:sparse-postorder-traversal} Starting from active primitives (orange nodes), sparse traversal follows active branches until the shadow condition is met. This triggers stacking of \emph{blob} bitfield for later use. Unstacking happens when the current node corresponds to the ancestor stored at the top of the stack. Numbers correspond to loop turn in Algorithm~\ref{alg:sparse-traversal}}.
\end{figure}

Thanks to the compactness property of our operators, we only need to process a subset of the blobtree nodes in order to evaluate a local field $f_\mathcal{A}$. This set of nodes, called  \emph{active nodes} are the nodes that contain at least one active primitive in their left or right subtree.

We rely on the access to the addresses $(P_i)_i$ of \emph{active primitives} provided by the A-buffer in order to propose a sparse bottom-up blobtree traversal algorithm that iterates only over the \emph{active nodes} instead of the whole tree.  
This traversal can then be used as the underlying algorithm for direct field evaluation or on-the-fly tree pruning.

\begin{algorithm}[tb]
\caption{ \label{alg:sparse-traversal} Sparse blobtree traversal. 
\\TYPE: floating-point for direct field evaluation, void for treepruning
\\$primitive$ and $operator$ are node processing methods } 	
    \begin{algorithmic}
        \STATE{$stack \leftarrow emptyStack<uint,TYPE>(STACK\_SIZE)$}
        
        \FOR{$i \in [0; n[ $} 
          \STATE { $N_{blob} \leftarrow B[4 P_i]$ }
          \STATE { $N_{data} \leftarrow primitive (N_{blob}, 4(P_i+1))$ }
          \STATE{ \#1 : Clamp $N_{blob}$ ancestor}

          \WHILE{ $ !(
      ((i+1<n \land  shadow(N_{blob},P_{i+1}))$ \\
       $\hskip\algorithmicindent \lor (i+1=n \land  (stack.empty() \land !valid(ancestor(N_{blob}))))
                       ) $ } 
 
              \STATE{$P_{op} \leftarrow ancestor(N_{blob})$}
              \STATE{$N_{blob}      \leftarrow B[4P_{op}]$}
              \IF{$!stack.empty()$}
                \STATE{$C_{blob} \leftarrow stack.top().blob$}
                \IF{$popRequired(P_{op}, C_{blob}, N_{blob})$}
                   \STATE{$C_{data} \leftarrow  stack.top().data$}
                   \STATE{$stack.pop()$}
                   \STATE{$N_{data} \leftarrow operator(C_{data}, N_{data}, N_{blob}, 4(P_{op}+1))$}
                \ENDIF
              \ENDIF
            
              \STATE{\#2 : Clamp $N_{blob}$ ancestor}

          \ENDWHILE
          
          \STATE{$stack.push(\{N_{blob}, N_{data} \})$} 
          \COMMENT{\#3}

        \ENDFOR
        
        \RETURN { stack.top().field }

	\end{algorithmic}
\end{algorithm}

\subsection{Parent pointer sparse traversal}

First, the post-order linearized tree layout of $B$ allows executing a stackless bottom-up post-order traversal by simply processing nodes in their memory storage order. 
This order guarantees that each node is processed after its two children subtrees (i.e., when intermediate field values for left and right subtrees are available). Therefore, our traversal needs to preserve this property to perform unique memory access to each active node in the tree.
Besides, traversing only active nodes can be done by going up iteratively from each active primitive $P_i$ following parent pointers until the tree root is reached. Doing so from more than one $P_i$ would result in several traversals of a subset of tree nodes, as shown in Figure~\ref{fig:shadowing} (left).
Therefore, each parent pointer traversal should be interrupted before reaching a node that cannot be evaluated from already processed nodes.

\paragraph{Shadowing condition:}
As shown in Figure~\ref{fig:shadowing} (middle), the first node of a subtree to be processed is always an active primitive. 
Until this primitive is processed, the parent of the subtree cannot be processed. This implies that a parent pointer should not be used to dereference a node if it points after the next unprocessed active primitive.
Given an unprocessed primitive address $P$ and a \emph{blob} bitfield $N$,
we define the \emph{shadowing condition} as:
\begin{equation}
    shadow(N,P) \rightarrow ancestor(N)>P \ ,
\end{equation}
which is illustrated in Figure~\ref{fig:shadowing} (right).

Following these principles, our traversal algorithm is provided in Algorithm~\ref{alg:sparse-traversal}, and Figure~\ref{fig:sparse-postorder-traversal} highlights the different phases of the algorithm.
First, stacking of intermediate state - i.e., shadowing \emph{blob} and additional data for field evaluation - happens at most once per primitive - see marker \#3 in the algorithm - when the shadowing condition is met, or the tree root has been reached.
Secondly, thanks to the behavior of operator $G_\cup$, a node with pointer $P_{op}$ only needs to be evaluated if both its operands are active nodes. In this case, the right operand corresponds to the currently processed tree branch $\{N_{blob},N_{data}\}$, and the left operand is on the top of the stack. This condition can be written as:
\begin{equation*} 
popRequiredA(P_{op}, C_{blob}, ...) \rightarrow P_{op} = ancestor(C_{blob})
\end{equation*}
with $C_{blob}$ the \emph{blob} bitfield stored at the top of the stack. 

As intermediate \emph{blob} need to be pushed on the stack, Algorithm~\ref{alg:sparse-traversal} can increase the stack memory required for direct field evaluation. This problem is avoided in our complete pipeline (see Section~\ref{sec:synchronized}).

\begin{wrapfigure}[11]{r}{3.5cm}
    \hspace{-5mm}
    \includegraphics[width=\linewidth]{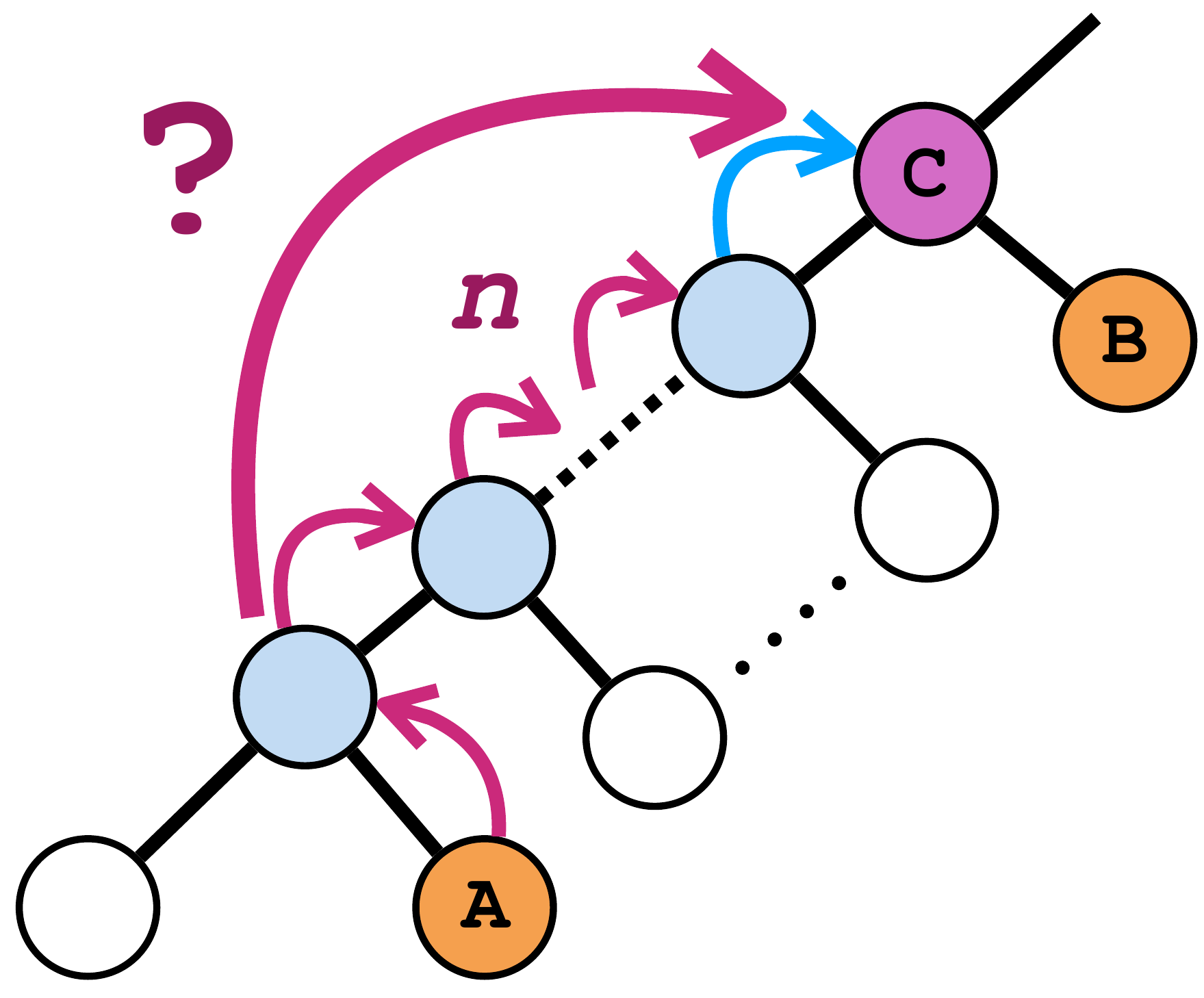}
\end{wrapfigure}
\paragraph{Left heavy tree} In order to limit field evaluation stack size, left-heavy trees should be favored~\cite{grasberger2016}; indeed, a comb-like tree, as depicted in Figure inset, only require a stack of size 1.
However, directly applying Algorithm~\ref{alg:sparse-traversal}, traversing such trees requires processing all operators stored after the first active primitive. This can represent most of the blobtree operators, skipping only unused primitives and requiring considerable work for large trees.
We propose replacing parent pointers with another type of ancestor pointer to solve this problem. 

\begin{figure*}[h]
    \centering
    \includegraphics[width=0.85\linewidth]{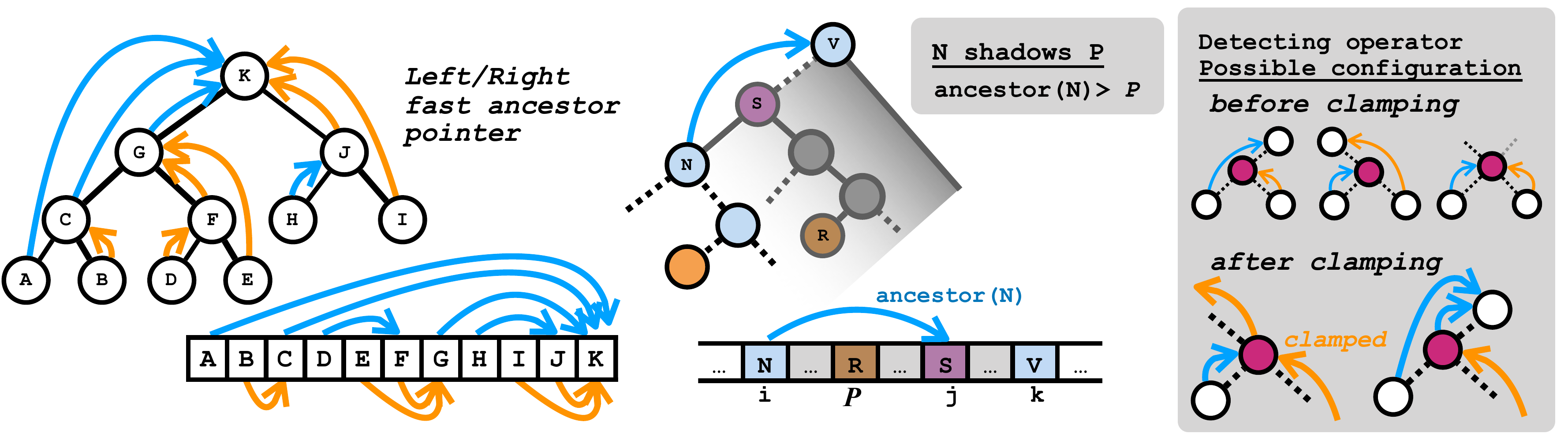}
    \caption{\label{fig:fast-ancestor}
    \emph{Left:} Fast left and right ancestor pointers in a blobtree.
    \emph{Middle:} Shadowing condition remains valid when using fast pointers.
    \emph{Right:} Possible configurations to detect node with two operands before and after clamping of parent pointer with the ancestor of the top \emph{blob} on the stack.
    }
\end{figure*}

\subsection{Fast ancestor pointer}

We propose to replace parent pointers with pointers providing longer jumps combined with adapted rules to detect active operators.
Given a node pointer $P$ and its parent pointer $A$, we define the fast ancestor pointers as :
\begin{equation*} 
    fastAncestor(P,A) \rightarrow  
    \left\{ 
    \begin{array}{ l }
        A, \text{if } isLeft(B[4P])=isLeft(B[4A]))\\
        fastAncestor(A, ancestor(B[4A])),   \text{otherwise}
    \end{array}
    \right.
\end{equation*}
As shown in Figure~\ref{fig:fast-ancestor}, fast pointers allow longer jumps provided that the jump direction - left or right -  does not change. They only depend on the tree structure and are precomputed. 

As fast pointers can only increase the ancestor pointer of a node, the shadowing condition remains unchanged.
However, the detection of the operators to process needs to be modified. Otherwise, operators might be missed (see configurations in Figure~\ref{fig:fast-ancestor}). This is solved by first clamping the fast pointer of current node $N_{blob}$ by the ancestor of the latest \emph{blob} in the stack - $ancestor(stack.top()$ if it exists - at marker \#1 and \#2 of the algorithm. 

After clamping, only two configurations can require the processing of an operator (see Figure~\ref{fig:fast-ancestor} bottom left), and the associated condition is the disjunction of previous \emph{popRequiredA} (blob at the top of the stack pointing to the current node)
with:  
\begin{align*}
        & popRequiredB(P_{op},  C_{blob}, N_{blob}) \rightarrow \\
& \qquad ancestor(N_{blob})\geq ancestor(C_{blob}) \\
& \qquad  \land ( !valid(ancestor(N_{blob})) \lor isLeft(N_{blob})) 
\end{align*}
which corresponds to bottom right configuration in Figure~\ref{fig:fast-ancestor}: the operator is a left child (second part of the new conjunction)
 and have an ancestor superior or equal to the one stored on top of the stack (first part).

%% file: core_synchronized.tex
\section{Synchronized tracing with prunned tree}
\label{sec:synchronized}

\begin{figure}[h]
    \centering
    \includegraphics[width=0.9\linewidth]{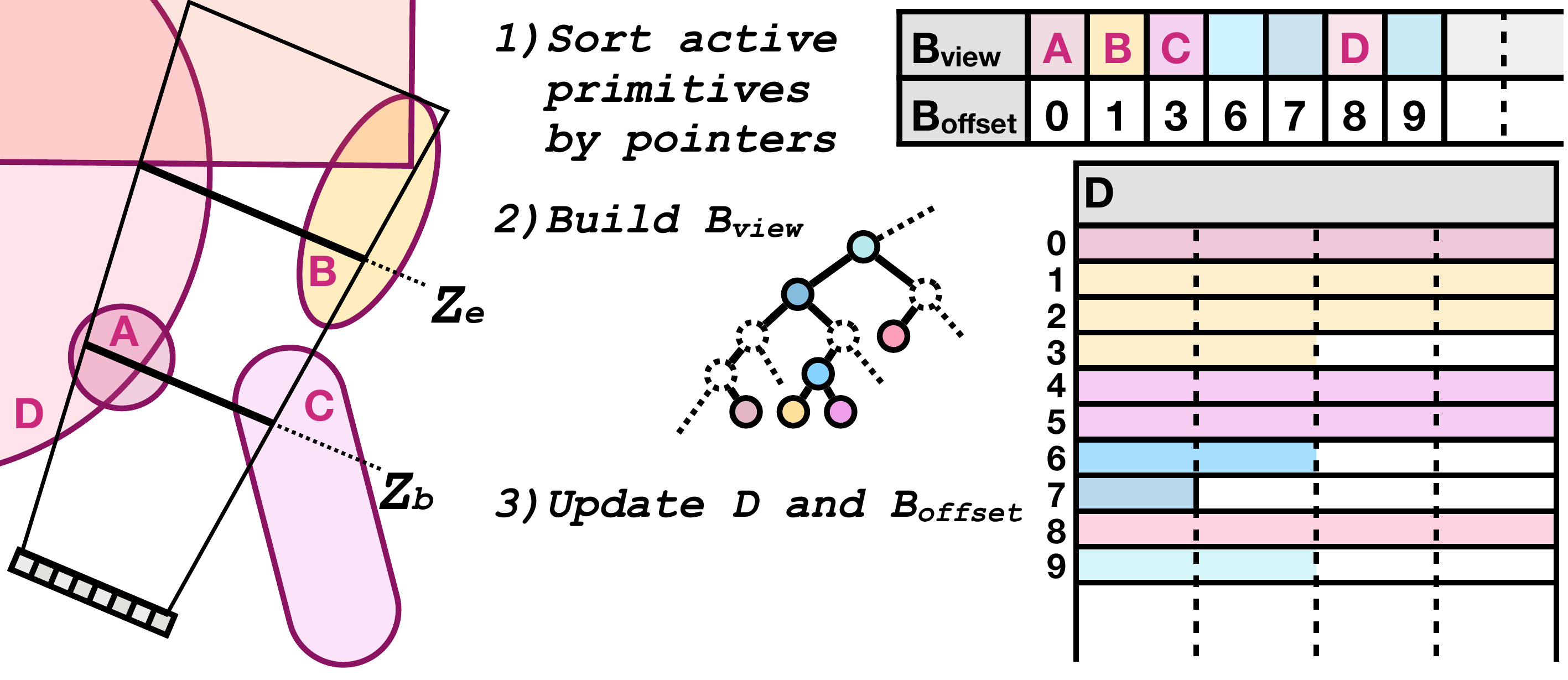}
    \caption{\label{fig:shared-blobtree-view} In order to process a subfrustum, we first reconstruct a prunned tree view from the list of active primitive using the sparse blobtree traversal. The tree view consist of a shared cache to store node parameters and two arrays storing 1) active node \emph{blob} bitfields, 2) offsets of node parameters into a local cache. 
}
\end{figure}
    
In order to maximize throughput, GPU architectures require coherent processing within workgroups. In our context, this means that all threads of a workgroup should evaluate the same functions when node interpreters are used during blobtree evaluation.
To achieve this objective, ray buckets are processed in a compute shader; each workgroup corresponds to a 8 by 8 screen tile. Ray processing is synchronized within a workgroup and relies on shared memory and on-the-fly tree pruning for faster field evaluation.

The main loop of the compute shader, as depicted in Algorithm~\ref{alg:ray_process}, consists of three main stages: tracking active primitives, construction of a pruned blobtree view, and processing of a ray interval using sphere-tracing. 

\begin{algorithm}[tb]
	\caption{ \label{alg:ray_process} ComputeShaderMainLoop(i,j)). } 	
  \begin{algorithmic}
  \STATE{$all\_found \leftarrow false, r_{found} \leftarrow false$}
  \STATE{shared: $root\_found[gl\_SubgroupID] \leftarrow false$}
  \STATE{shared: $z_{begin}, z_{end} \leftarrow  0, i \leftarrow  0$}
	\WHILE{$!all\_found \land i<n_{prim}$} 
    	\IF{$gl\_SubgroupID=0$}
            \IF{glSubgroupElect()}
              \STATE{$[z_{begin},z_{end}], i \leftarrow fetchNewPrimitives()$} 
              \STATE{$buildTreeView()$} 
            \ENDIF
		    \STATE{$synchronize()$}
		    \STATE{$fetchNodeData()$} 
		 \ENDIF
     \STATE{$synchronize()$}

     \IF{$!r_{found}$}
         \STATE{$[t_0,t_1] \leftarrow ray(i,j).fromNDC([z_{begin},z_{end}])$} 
         \STATE{$r_{found} \leftarrow processInterval( ray(i,j),[t_0,t_1])$} 
         \STATE{$s\_root\_found[gl\_SubgroupID] \leftarrow subgroupAnd(r_{found})$} 
     \ENDIF
     \STATE{$synchronize()$}
     \STATE{$all\_found \leftarrow \bigwedge_k root\_found[k]$}      
	\ENDWHILE
	\end{algorithmic}
\end{algorithm}

\paragraph{Avoiding thread divergence}

Workgroup synchronization relies on subfrustum processing. At each turn of our main processing loop, each thread processes a ray interval that is defined by a shared depth range $[z_{begin},z_{end}]$ for the whole workgroup. This range is defined in normalized device coordinates (NDC). The uniqueness of the depth range combined with the tile-based nature of our A-buffer guarantees that all threads will use the exact same set of active primitives during the interval processing.

\paragraph*{Tracking active primitives}
We store the active primitives in shared memory. Let us define $z_{b,j}$ and $z_{e,j}$ as the entry and exit depth - defined in NDC - inside the $i$-th primitive of interest associated with the current screen tile.

At the beginning of each loop turn, we first discard all active primitives verifying $z_{e,j}\leq z_{end}$. 
We then define the next beginning of subfrustum to process as: 
$$z_{begin} = \max(z_{end},z_{b,i+1})$$
with $i$, the index of the last fetched primitive (i.e., we avoid processing of empty space when possible). 
Next, we fetch new primitives of interest and compute at the same time the next depth range to process using a simple heuristic. Primitives are fetched until the conjunction of the four following conditions is false:
\begin{itemize}
  \item $z_{b,i+1} <= \max_{j<i} z_{e,j}$  \# no empty space should be processed
  \item $z_{b,i+1} - z_{begin}< d$ \# limit variation of number of primitive in depth range
  \item \# no more than 6 new primtives
  \item $n_{active}<M_{overlap}$ \# limited number of active primitives
\end{itemize}
with $n_{active}$ the number of active primitives, $d$ an arbitrary value set to ($1:20$ of the scene depth), and $M_{overlap}$ an implementation limit on the number of active primitives that can be stored. The last condition is essential to limit shared memory usage.

To avoid processing of empty space, the end of the subfrustum is then defined as:
$$z_{end} = \min\left(z_{b,i+1}, \max_{j<i} z_{e,j}\right) \ .$$

As discussed in Section~\ref{sec:bottom_up_eval}, the list of primitives needs to be sorted by node address in the blobtree $B$. 
For simplicity, this process is performed by a single thread, and a simple insertion sort is used to keep primitive sorted.

\begin{algorithm}[tb]
	\caption{ \label{alg:prunned_tree_eval} Prunned tree evaluation. } 	
    \begin{algorithmic}
    \STATE{$stack \leftarrow emptyStack<float>(STACK\_SIZE)$}
    \FOR{$i \in [0; 2num\_prim-1[ $} 
        \STATE{$nodeop \leftarrow nodeType(B_{view}[i])$}
        \STATE{$offset \leftarrow B_{offset}[i]$}
        \IF{$isPrimitive(B_{view}[i])$} 
          \STATE{$f_{node} \leftarrow evalPrimitive(nodeop, offset, ...)$} 
        \ELSE
          \STATE{$f_{right} \leftarrow stack.top()$}
          \STATE{$stack.pop()$}
          \STATE{$f_{left} \leftarrow stack.top()$}
          \STATE{$stack.pop()$}
          \STATE{$f_{node} \leftarrow evalOperator(nodeop, offset, f_{left}, f_{right}, ...)$} 
        \ENDIF
        \STATE{$stack.push(f_{node})$} 
  \ENDFOR
	\RETURN{stack.top()}
	\end{algorithmic}
\end{algorithm}

\paragraph{Local prunned tree view}

The processing of a given subinterval is likely to require more than one field evaluation, especially for grazing rays. Instead of using the sparse tree traversal for direct field evaluation, we use it to reconstruct a local pruned tree view in shared memory.
This provides several benefits: the field evaluation loop is highly simplified, no direct access to $B$ is done during field evaluation (decreasing the danger of cache eviction), and only nodes with two operands are stored then used during field evaluation. The simplified field evaluation is a standard bottom-up traversal and is provided in Algorithm~\ref{alg:prunned_tree_eval} for reference. 

We organize the pruned tree view in three main arrays (see Figure~\ref{fig:shared-blobtree-view}). The largest - designated as $D_{cache}$ -  is used to store active node parameters. The two others have size $2M_{overlap}-1$ and store per node information: we store \emph{blob} bitfields in the array $B_{view}$ and use array $B_{offset}$ to save the index of associated node parameters in $D_{cache}$.

The pruned tree construction is done after the active primitives' update; it is done in two main stages. First, a single thread creates the pruned blobtree view $B_{view}$ using the sparse traversal to store \emph{blob}s (storing them in traversal order to preserve their relative order). Then, the parameter cache $D_{cache}$ is filled collaboratively. Offsets in $B_{offset}$ are updated accordingly. For each \emph{blob} bitfield, node parameters to be read are distributed between threads of a subgroup until all threads have an address, then parallel read and write are used to transfer memory. This is repeated until all \emph{blob} of active primitives are processed, or $D_{cache}$ is full. In this latter case, node parameters are kept in main memory, and the node address in $B$ is recorded in $B_{offset}$.

The tree pruning only requires storing the \emph{blob} on the stack, and the field evaluation the intermediate field values. Because both stacks are not used at the same time, the effective memory requirement is decreased and matches the one of a regular bottom-up traversal.

%% file: core_interdiff.tex
\section{Intersections and differences}
\label{sec:interdiff}

Intersections and differences cannot be handled the same way as union operators. In order to compute the full expression correctly, intersection operators should never be bypassed: fast ancestor pointers are cut at the level of intersection operators. Similarly, right subtree parent pointers should not bypass difference operators.
Indeed, if such nodes are bypassed by their right operands (or left for intersection), the operator's behavior would be changed to a union-like node (returning the minima of its operands). This also means that intermediate intersection and difference with a single operand should not automatically propagate the child field value to their parent (union behavior) 

\paragraph{Evaluate operators outside of child ROIs:}  Let us consider what happens for a smooth intersection operator verifying Equation~(\ref{eq:smooth_comp}) - for differences, similar reasoning can be followed for the right operand; the case of the left operand is similar to the union.
If both operands are active, the operator can be evaluated as usual. Let us assume we have $f_1 \notin R_{c_1}$ without loss of generality if at least one operand is outside of its range of interest. Then we have, $op(f_1,f_2,k) = max(f_1,f_2)$ and two cases can arise:
\begin{itemize}
    \item $f_1>f_2$, then we have $op(f_1,f_2) = f_1 > d_{c_1} \geq d_p$, which mean that the result of the operator will not be required by the parent and an infinite value can be returned safely or an ignore status can be propagated in the tree.
    \item $f_1 \leq f_2$, then we also have $f_2 \notin R_{c_2}$ and the same reasoning can apply. 
\end{itemize} 
Therefore, if only one intersection operand is active, it should be ignored by returning an infinite value outside parent ROIs. The same behavior should be used for the right operand of difference nodes if alone.

While this strategy can be used to define $f_\mathcal{A}$ during direct field evaluation,
it cannot be used with on-the-fly tree pruning, where only active nodes with two active operands are stored.

 We define the behavior that a node should follow in the absence of one of its operands in the \emph{ignoreMod} field of the \emph{blob} of the node; possible values are:
 \begin{itemize}
    \item  $never\_ignore = 00$, i.e. union-like operator
    \item  $ignore\_if\_right\_absent = 01$ 
    \item  $ignore\_if\_left\_absent = 10$, i.e. difference-like operator
    \item  $ignore\_if\_any\_absent = 11$, i.e. intersection-like operator
\end{itemize}

Let us define the \emph{usage} status of a node as a bit with a value of 1 if the node is to be taken into account during field evaluation, and 0 otherwise. We modify the \emph{nodeop} of traversed nodes in order to take into account the \emph{usage} status of their operands. Our operator interpreter has three reserved values for this purpose:
\begin{itemize}
\item  $case\ 0 :\ return\  \infty$ // branch is unused
\item  $case\ 1 :\ return\ right\_field$
\item  $case\ 2 :\ return\ left\_field$
\end{itemize}

We compute the \emph{usage} status during blobtree traversal.
A \emph{usage} stack - represented as a bitfield for compactness - is used to track \emph{usage} status of nodes (all active primitives start with a value of 1).
In order to compute the \emph{usage} status of a node operator with pointer $p$, we start by retrieving its two operands \emph{usage} status from the stack and store them in a bitfield $b_{children}$.
Then, we compute a candidate $nodeop$: 
$$op\_type \leftarrow  ((~b_{children} \& ignoreMod(blob)) == 0) ? b_{children} : 0$$
If all bits of $op\_type$ are set, the node \emph{blob} should not be modified; else, it is used to replace the \emph{nodeop} of the node \emph{blob} in the pruned tree view.
Similarly, the \emph{usage} status of the node is defined as: 
$$b_{children}=0 \lor (~b_{children} \& ignoreMod(blob)) > 0)$$
and is pushed to stack for later use. If the \emph{usage} status of the root is 0, there is no 0 iso-crossing in the studied volume.

%% file: implementation.tex
\section{Implementation details}
\label{sec:implem}

Our pipeline was implemented in C++ using the OpenGL library (version 4.4) and shaders were programmed in GLSL using extensions for subgroup instructions. 
We present here the default parameters used for most of our tests and their implications on our rendering pipeline.

\paragraph{Approximate SDF}
We use the following primitives for our examples: spheres, ellipses, torus, cubes, sphere-cones, and quadrics. Base primitives can be transformed with translation and rotation. Our set of operators consists of standard CSG operators (based on $\min$ and $\max$ functions), smooth polynomial operators presented in section~\ref{sec:taxonomy_blend}, and procedural nodes. 
Remember that we can use at most 32 primitives and 32 operators at once; if more are required, a possible solution would be to have a specialized node performing a supplementary switch statement based on node data.

\paragraph{Ray processing} 
Subinterval along rays are processed using a variant of overrelaxed sphere tracing. The first non-intersecting sphere is stored for later reuse, and overrelaxation is deactivated until the saved sphere is reached. We use an overrelaxation factor of $1.7$.

We set the Lipschitz bound to obtain artifact-free rendering on all presented examples. In practice, we set it to 1.45 to accommodate for field compression associated with our compact operators. 
Replacing sphere-tracing with segment-tracing would provide a more robust solution.

We do not limit the maximal number of steps as usually done in sphere-tracing implementation but only set a minimal step size equal to $0.005$. For all our examples, this means the step size is around $1:32000$ of the scene bounding box maximal dimension.

Normals are computed as iso-crossing depth differentials.

\paragraph{Memory allocation}
 
Memory allocation - registers and shared memory - in shaders play a considerable role in the GPU occupancy and, therefore on performance. In our implementation, register allocation is impacted by two central aspects: the set of primitives and operators type and field evaluation stack. 
In practice, we limit the stack to $22$ entries, limiting the maximal number of primitives in a perfectly balanced tree to 4 million. Conversely, processing of \emph{left} comb trees is not limited by stack size, and a right-heavy tree can be rebalanced to obtain a left-heavy one (an optimization we still need to put in place). Furthermore, should a processing method with larger stack requirements be used - e.g., segment tracing - stack size can be halved while still being above the requirement of all tested examples.

The two main parameters impacting shared memory usage are the maximal number of overlap between primitives - the size of $B_{view}$ and $B_{offset}$ is twice this value - and the maximum amount of primitive and operator data stored in the node data shared cache.
We set the maximal number of primitive overlap to 96 and the maximal cache size to 3072 bytes (additional node parameters are read normaly when going above this limit). 
If the maximal number of overlaps is reached, this results in an average of 8 floats per node (usually, operators require fewer data than primitives).

For the A-buffer, we reallocate memory on demand. At a resolution of $1024 \times 1024$, storing 200 entries per pixel requires only 52.4MB - against the 3.36GB required at full resolution - well below GPU memory resources.

%% file: discussion.tex
\newcommand{\MagicianTroll}{Fig.\ref{fig:teaser}}

\input{Z_render_user_created.tex}

\section{Results}
\label{sec:discussion}

We tested our implementation on two nVidia graphic cards presenting a large difference in number of CUDA cores, namely an nVidia Quadro P1000 (with 512 cores and 4 GB memory) and an nVidia GeForce 2080 RTX (with 4352 cores and 11 GB memory).
The Quadro P1000 runs on a Ubuntu laptop with an Intel Core i-7 8850H (2.6GHz, 6 core), and the 2080 RTX runs on a Window workstation with an Intel Core i7-4770K (3.5GHz, 4 cores).

We first provide general runtime and memory usage statistics on complex implicit models. 
We then compare our method to recent state-of-the-art rendering technique for implicit surface, namely~\cite{Keeter2020}.


\paragraph{GPU runtime statistics} 
We rely on deferred shading, provided rendering time only accounts for the construction of the G-buffer (e.g., the iso-surface extraction and not the shading itself). We always record the average rendering time over 20 frames from a given viewpoint. Runtimes for the 2080 RTX were measured at a resolution of $1720 \times 1376$ (resolution used for all figures), the ones for the Quadro P1000 were measured at $1024 \times 1024$.

We provide statistics for several user-created examples consisting of hundreds of primitives (Figure~\ref{fig:teaser} and~\ref{fig:examples}).
Runtimes and preprocessing times are provided in Table~\ref{tab:runtime_table}. Runtimes range from interactivity with the Quadro P1000 (more than 30fps) to real-time with the 2080 RTX (more than 60fps). 
Preprocessing includes volume of interest generation (with the range of interest propagation) and GPU blobtree initialization (with fast pointers computation). However, this step is not required when only primitives are modified. 
Our monothreaded implementation does not impair interactivity for blobtrees consisting of a few hundred primitives. 
For interactive editing of larger blobtrees, a strategy to re-use part of the preprocessing would be required.

Figures~\ref{fig:teaser} and~\ref{fig:examples} also show that per-tile maximal shared memory usage (maximal number of primitives overlap and node data used) remains well below allocated shared memory resources (described in Section~\ref{sec:implem}).

We also provide more complex synthetic procedural examples (see Figure~\ref{fig:synth-comp}), which will be used for comparison to~\cite{Keeter2020}. All but one of the synthetic examples are generated by a uniform distribution of procedural objects consisting of a few primitives. We only use the type of primitive present in~\cite{libfive}.
The last example is using multiresolution instancing. 
Subobjects are created either using smooth or sharp operators, and subobjects are combined together with a left heavy tree of sharp union operators.
Runtime and preprocessing time for synthetic examples are provided in Table~\ref{tab:preprocess_comparison_keeter} and~\ref{tab:runtime_comparison_keeter}. For the 2080 RTX, they remain above 60fps for most examples. 

For both user-created and synthetic examples, runtimes with the 2080 RTX are around $5-9\times$ faster than with the Quadro P1000 showing a good scaling with the number of GPU cores as long as the maximal number of primitive overlaps in the scene remains below implementation limits (i.e., 96 overlaps for provided runtimes).


\input{blend_graph.tex}
\subsection{Analysis}

In order to provide some insight on the impact of each of our contribution, we propose runtime comparisons between variants of our pipeline (see Table~\ref{tab:runtime_table}). We can see that all aspects are key to decrease runtime by a factor 10 compared to non-synchronized processing performing sparse tree traversal without fast pointers relying on a full-resolution A-buffer.

\paragraph{Positioning with respect to~\cite{grasberger2016}}
Our method is built on the work of~\cite{grasberger2016}. The latter uses a spatial acceleration structure that needs to be updated when primitive parameters change (typically happening during modeling). On our side, preprocessing is only required if the tree itself is updated and consists of a simple tree traversal with linear complexity independent of primitive spatial configuration.
Rendering-wise, the spatial embedding of~\citet{grasberger2016} acceleration structures would make it more difficult to traverse ray in a coherent way (a given screen tile could span multiple space subdivisions at once).

\paragraph{Large A-Buffer} In presence of large number of small primitives with large blend range as in Figure~\ref{fig:synth-comp} (R4C3-4), the building time and memory consumption as well as local number of primitive overlap can increase drastically. In such configuration, building the A-buffer become the bottleneck of the pipeline. When only union-like operators are used, this could be alleviated with an occluder depth path, rendering primitives without blend~\cite{Bruckner2019}. 
Reducing the blend range allow to render properly the example R4C3 as shown if Figure~\ref{fig:zoom_multi_res}.

\paragraph{Impact of tree structure} We have checked the impact of the tree layout on runtime; for Figure~\ref{fig:synth-comp} (R2C3), it varies by less than  $3\%$ between left heavy tree and fully balanced tree for both sharp and smooth union. 
As already discussed, the tree structure also has an impact in terms of stack size requirement and the left heavy tree should be favored. Depending on node types, rotation and child swaps can be performed to reduce stack requirements without impacting the field definition~\cite{grasberger2016} (a right heavy tree can be transformed into a left heavy tree to reduce stack requirement from linear to constant). However, we did not put this optimization in place.
Because of the range of interest propagation, adding an operator with a limited radius of influence at the top of the tree can provide faster runtime. In terms of modeling, this means primitives representing details should not be below operators used to blend a coarse shape.

\input{runtime_table-examples.tex}

\subsection{Comparison to~\cite{Keeter2020}}

Before discussing the runtime comparison with~\cite{Keeter2020}, it is essential to remember that both methods study different tradeoffs and that the overlap between types of shapes that can be processed is only partial. For instance, the approach proposed by~\cite{Keeter2020} can process any field functions as long as it is defined as arithmetic operations; this excludes procedural operations based on control flow (if, for, ...). On the contrary, our approach can support the latter, but we require the field function to be an A-SDF. 

In synthetic examples used for comparison (see Figure~\ref{fig:synth-comp}), all primitives but spheres have their own rotation. Such configuration is likely to happen in a free-form shape modeling context.
We test both sharp union and smooth union. For the latter, our implementation relies on our compacted operators (Equation~(\ref{eq:smooth_comp})) while we use standard operator (Equation~(\ref{eq:smooth-union}) for~\cite{Keeter2020} as our operator implementation rely on both an $if$ and an $isnan$ function. 

For the comparison, we used the available CUDA implementation of~\cite{Keeter2020} available at~\cite{mpr}. Comparisons are made using the nVidia Quadro P1000 on the Unix system.

\paragraph{Preprocessing}
For a small number of primitives combined with union operators - which can already represent a large number of arithmetic clauses - we observe similar preprocessing times that do not prevent interactive editing (see Table~\ref{tab:preprocess_comparison_keeter}).
For~\cite{Keeter2020}, observed preprocessing time starts to impair interactivity (more than 50ms) when the number of primitives goes above a few hundred, and it worsens when smooth unions are used. Our implementation remains below 50ms up to a few thousand primitives.
Our method requires between two and three times less memory to store the expression data on GPU (below 1MB for all examples but the multiresolution one), however, studied examples are not well suited for the subexpression re-use property of~\cite{Keeter2020}. 
It is important to remember that this preprocessing step is not a core part of the contribution of~\cite{Keeter2020} and our method and that it could be subject to optimization in both cases. Furthermore, provided adequate data management, both methods have in common the ability to change primitive parameters without requiring this preprocessing, a property well adapted for modeling tasks. 

\paragraph{Sharp union operator} Runtime comparisons are presented in Table~\ref{tab:runtime_comparison_keeter}. 
For examples from rows 0 to 2 included, generated with only union operator, we obtain rendering time reduction starting from $-62.0\%$ on the most straightforward configurations. The difference then increases with the number of the instanced subobject in the scene: contrary to~\cite{Keeter2020}, we never have to process the complete evaluation tree during rendering.
The multiscale examples from row 4 provide more significant improvement (from $-89.0\%$) due to the large number of small primitives in the scene.
Note that this is among the most favorable case for our method as the union operator allows computing thight volumes of interest, limiting the number of primitive overlaps and maximizing empty space skipping. 

\paragraph{Sharp intersection operator} In the presence of intersection operators discarding the majority of the primitive volumes (examples from row 3 using intersections between spheres), \cite{Keeter2020} provides rendering time reduction up to $-60.0\%$ due to more efficient space discarding in those configurations. However, when the number of primitives increases too much, our methods provide comparable performance (i.e., at 1536 primitives corresponding to 64 subobject instances).

Note that using specialized nodes to represent intersections between two primitives can provide significant runtime improvements. Using such approach, our method provides rendering time reduction from $-35.0\%$ to $-76.3\%$. However, such optimization can only be done in a specific context.

\paragraph{Smooth operator} When using smooth operators, large examples could not be rendered with~\cite{Keeter2020}.
On supported examples, smooth union produces a slightly higher time reduction for our methods  (from $[-62.0\%;-82.3\%]$ to $[-67.7\%;-85.9\%]$), always remaining under runtime off~\cite{Keeter2020} with a sharp union. For smooth intersection examples, runtime reduction is now in the range $[-18.7\%;-58.3\%]$ in favor of our method.
The importance of compactified blend is emphasized in Figure~\ref{fig:blend_prim}.

\subsection{Limitations}
Our rendering pipeline and its implementation present some limitations. Let us rapidly remind already discussed ones: we only support a limited number of primitive and operator types at once, a preprocessing is still required when the blobtree structure is modified, on the fly tree pruning is done by a single thread per workgroup, A-buffer construction time can increase significantly when a large number of small primitives are combined with large blend range, and finally, we assume a maximal number of primitive volume of interest overlaps in processed depth range.
When using smooth operators, this property is also a bottleneck for most existing techniques as the field evaluation complexity grows linearly with the number of overlaps. 

Apart from the last drawback, those limitations lie in our implementation rather than in the core of the method. 
Some strategies exist to alleviate them: for instance, partial results of tree preprocessing could be stored to drastically reduce preprocessing cost during the update of larger blobtrees, and processing by depth slabs can reduce A-buffer building time. 

\input{Z_comp_synth.tex}

\input{dexel_graph.tex}


%% file: Z_render_user_created.tex
\begin{figure*}[h]
    \centering
    \includegraphics[width=0.95\linewidth]{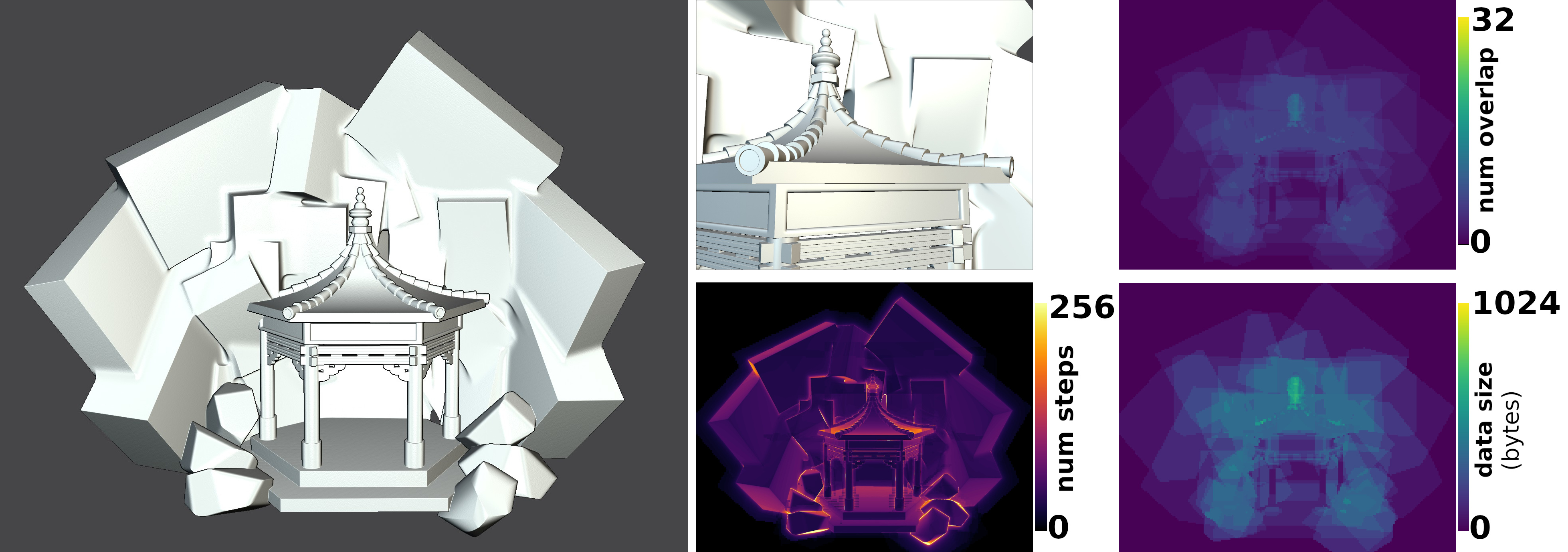}

    \caption{\label{fig:examples}  \emph{Left:} Render of an A-SDF model consisting of 159 primitives assembled using smooth and sharp unions, intersections and differences.
    \emph{Middle:} Close-up view (up) and number of field evaluations per ray (down - clampled for readability), large number of steps can be used locally without impacting interactivity. 
    \emph{Right:} Number of overlaps and size of node parameter cache $D$ per rendering tile, both remains well below implementation limits.
    }
  \end{figure*}

%% file: blend_graph.tex

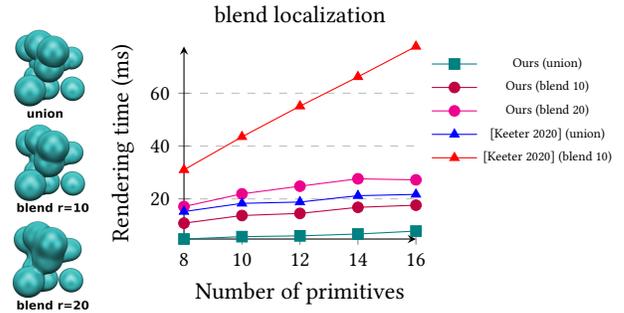
\begin{figure}[tbhp]
\includegraphics[width=0.15\linewidth]{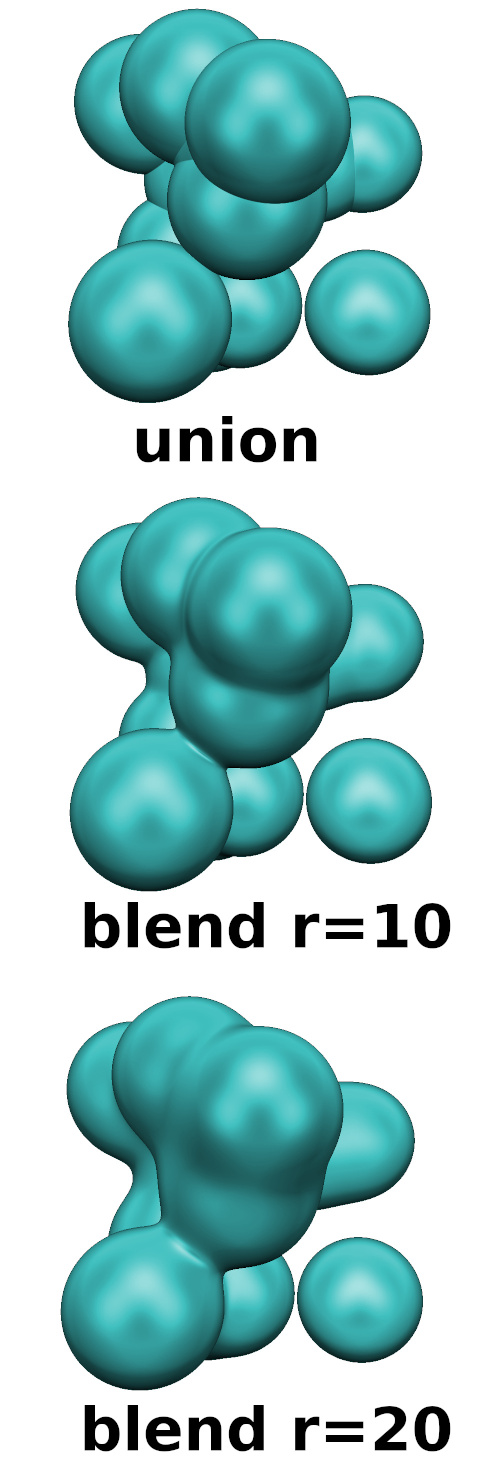}
\begin{tikzpicture}
\begin{axis}[
     axis x line=bottom,
     axis y line=left,
    scale=0.45,
    title={blend localization},
    xlabel={Number of primitives},
    ylabel={Rendering time (ms)},
    xtick={8,10,12,14,16},
    legend style={draw=none,font=\tiny},
    xticklabel style = {font=\small},
    yticklabel style = {font=\small},
    /pgf/number format/1000 sep={},
    legend pos=outer north east,
    ymajorgrids=true,
    grid style=dashed,
]

\addplot[color=teal, mark=square*,]
coordinates {
    (8,4.8) (10,5.7) (12,6.0) (14,6.7) (16,7.8)  }; 
\addplot[color=purple, mark=*,]
    coordinates {
(8,10.8) (10,13.7) (12,14.5) (14,16.8) (16,17.6)  }; 
\addplot[color=magenta, mark=*,]
    coordinates {
        (8,17.1) (10,21.9) (12,24.8) (14,27.6) (16,27.2)  }; 
\addplot[color=blue, mark=triangle*,]
    coordinates {
        (8,15.2) (10,18.3) (12,18.8) (14,21.2) (16,21.7)  }; 
\addplot[color=red, mark=triangle*,]
    coordinates {
        (8,31.0) (10,43.5) (12,55.1) (14,66.2) (16,77.7)  }; 
\legend{Ours (union), Ours (blend 10), Ours (blend 20), \cite{Keeter2020} (union), \cite{Keeter2020} (blend 10)}  
\end{axis}
\end{tikzpicture}

\caption{\label{fig:blend_prim} \emph{Left:} Renders of 16 spheres showing the three operator ranges used for comparison.
\emph{Right:} Runtime measured while progressively adding new primitives in the scene. Using classical smooth bounded blending, runtime scales linearly with the number of primitives in the scene. With our compact operators, we observe similar runtime as the one of~\cite{Keeter2020} using sharp CSG union (minimum function), which emphasizes the importance of ROIs.
}

\end{figure}

%% file: runtime_table-examples.tex
\begin{table*}[tbhp]
	\begin{center}
		\begin{tabular}{|l||c|c||c|c|c||c|c|c|  }
			\hline
			& & preprocess &  \multicolumn{3}{|c||}{nVidia Quadro P1000} & \multicolumn{3}{|c|}{nVidia GeForce 2080 RTX} \\
			\cline{4-9}
			\cline{4-9}
			Object &  \# of primitives & time & ParentPtr/Full  & FastPtr/Full & Synchronized  & ParentPtr/Full  & FastPtr/Full & Synchronized \\
			\hline
			\hline
			Astronaut (Fig.\ref{fig:teaser}) & 142 & 6 & 658 & 301 & 22.1 & 57.8 & 30.5 & 7.3 \\
			Temple (Fig.\ref{fig:examples})  & 159 & 7 & 1894 & 1028 & 35.6 & 124 & 81.5 & 7.9 \\ 

			\hline
		\end{tabular}
				\caption{Rendering and preprocessing times (in milliseconds). Rendering times are averaged over 20 frames with a resolution of $1024\times1024$ for the Quadro P1000 and $1720\times1376$ for the RTX 2080. \emph{Full/ParentPtr} and \emph{Full/FastPtr} correspond to regular processing with a full-resolution dexel buffer using respectively the sparse bottom-up parent pointer and fast ancestor pointer traversals. \emph{Synchronized} correspond to our full pipeline. 
				}
		\label{tab:runtime_table}
	\end{center}
\end{table*}


%% file: Z_comp_synth.tex
\begin{figure}[h]
    \centering

    \includegraphics[width=0.24\linewidth]{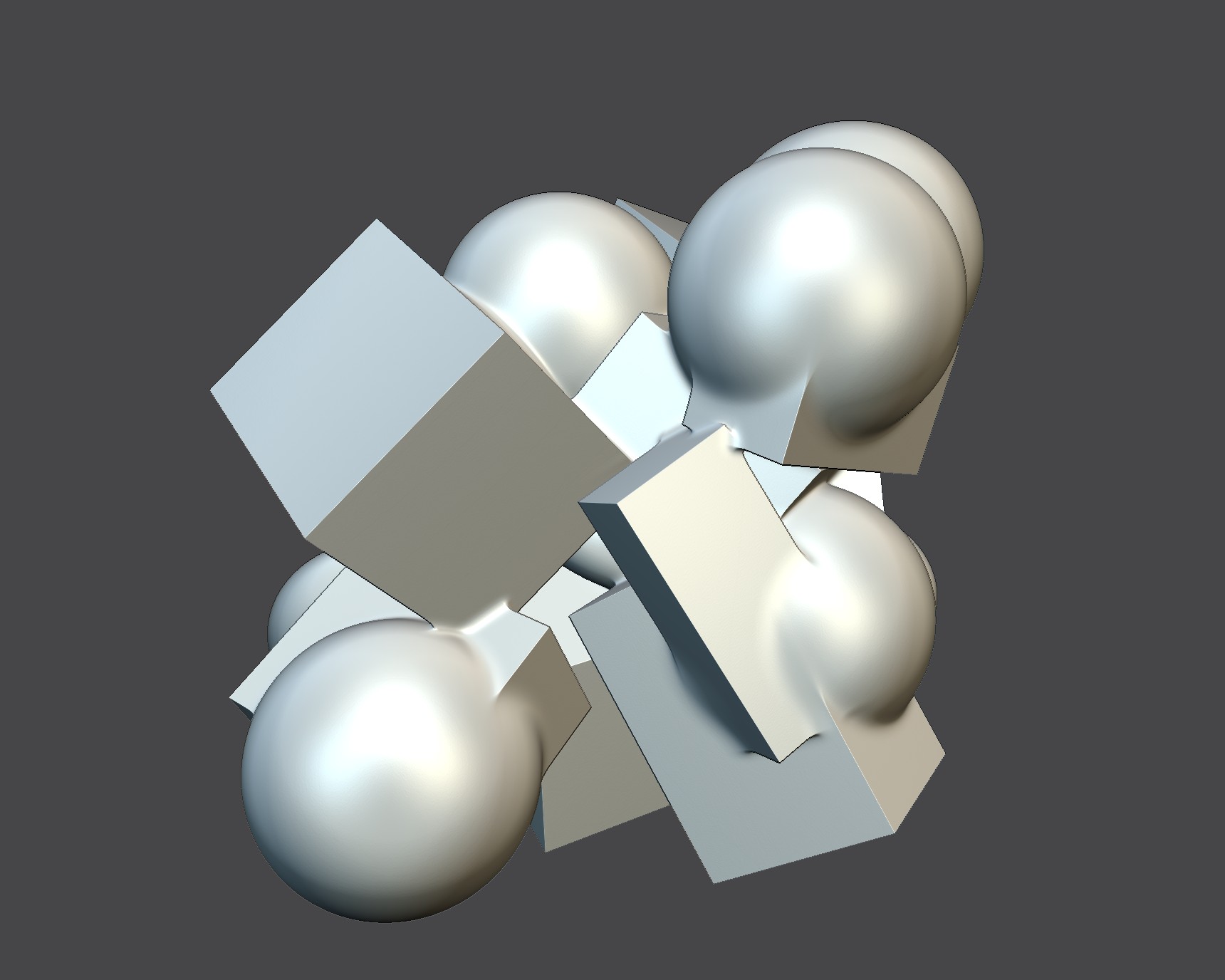}
    \includegraphics[width=0.24\linewidth]{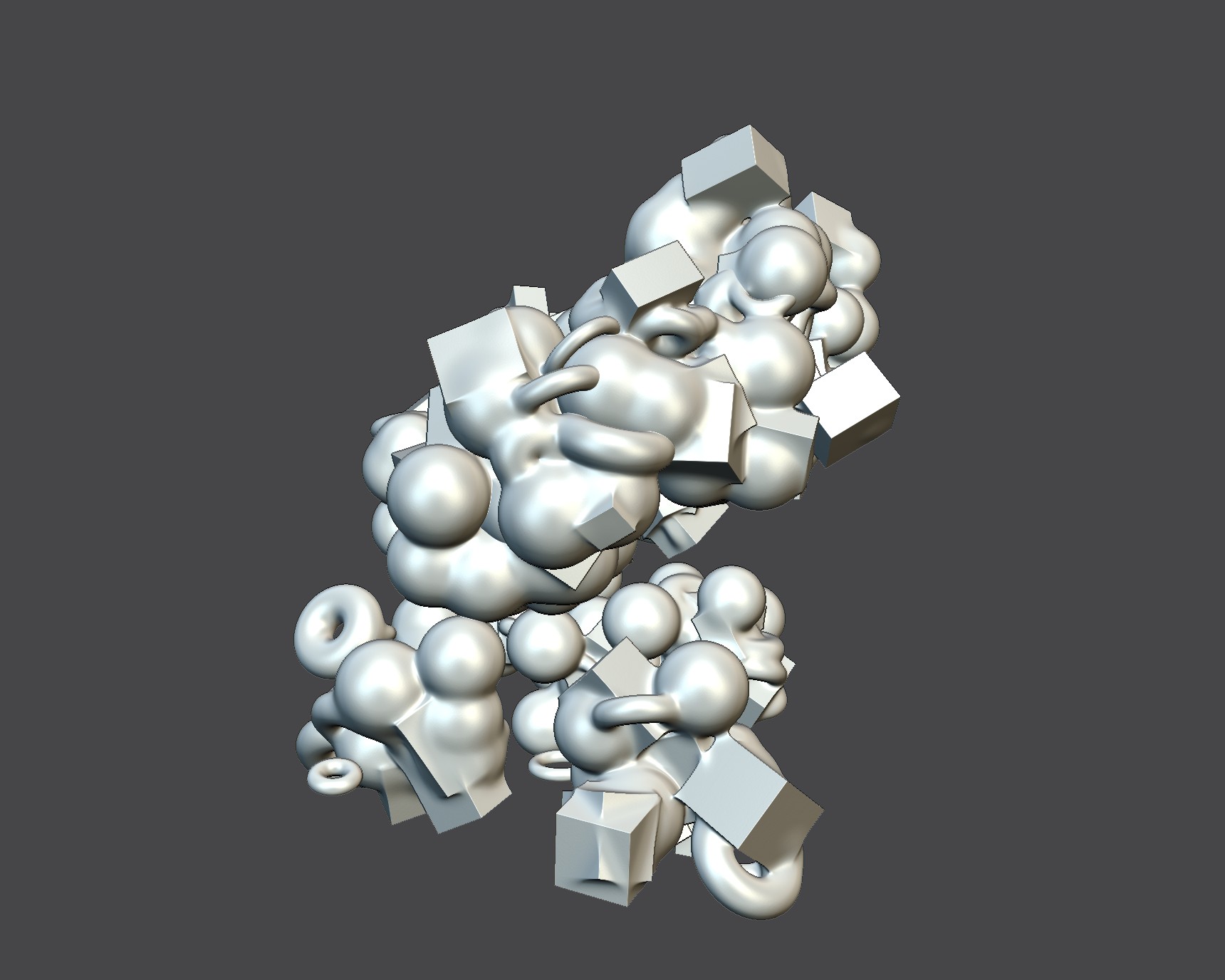}
    \includegraphics[width=0.24\linewidth]{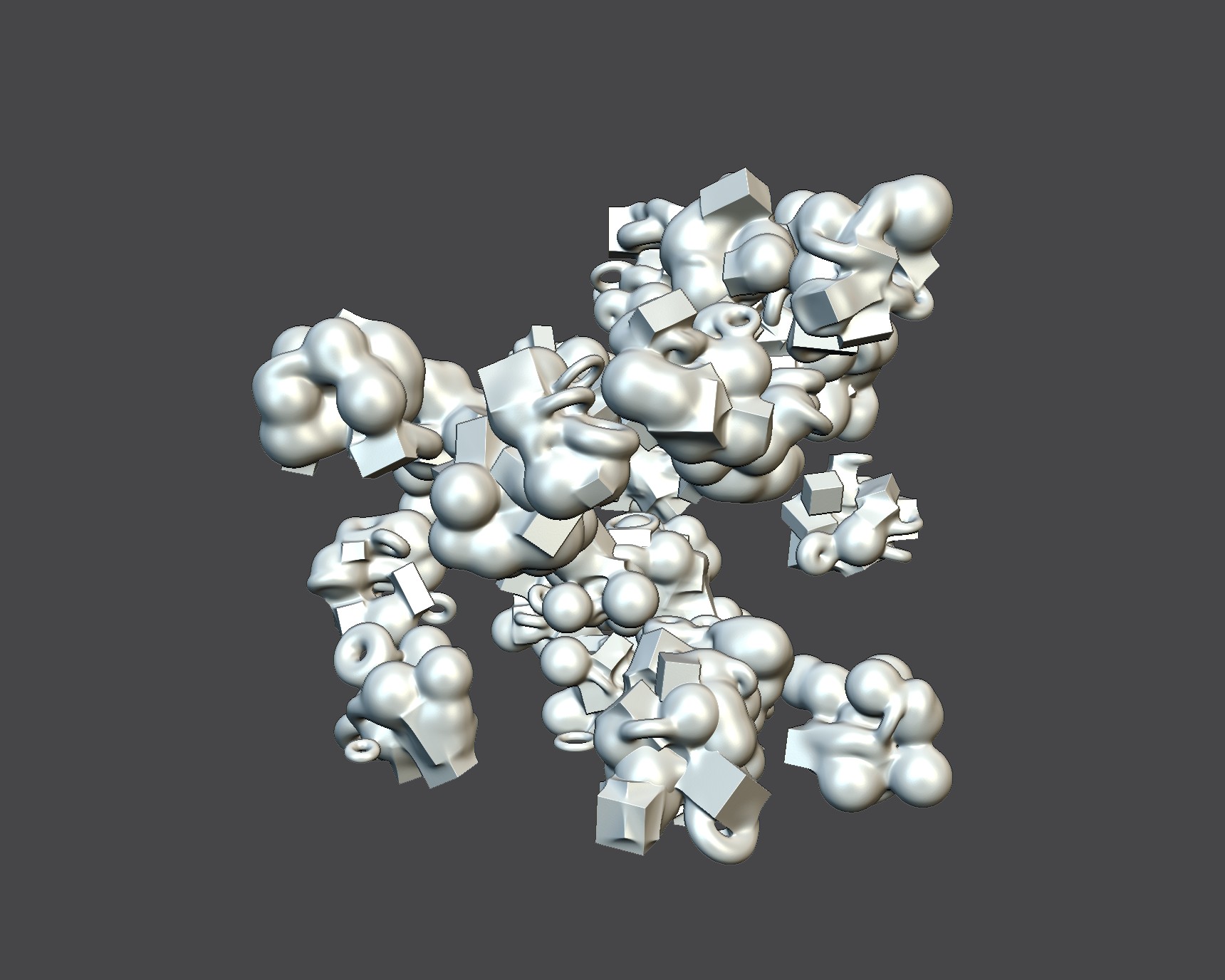}
    \includegraphics[width=0.24\linewidth]{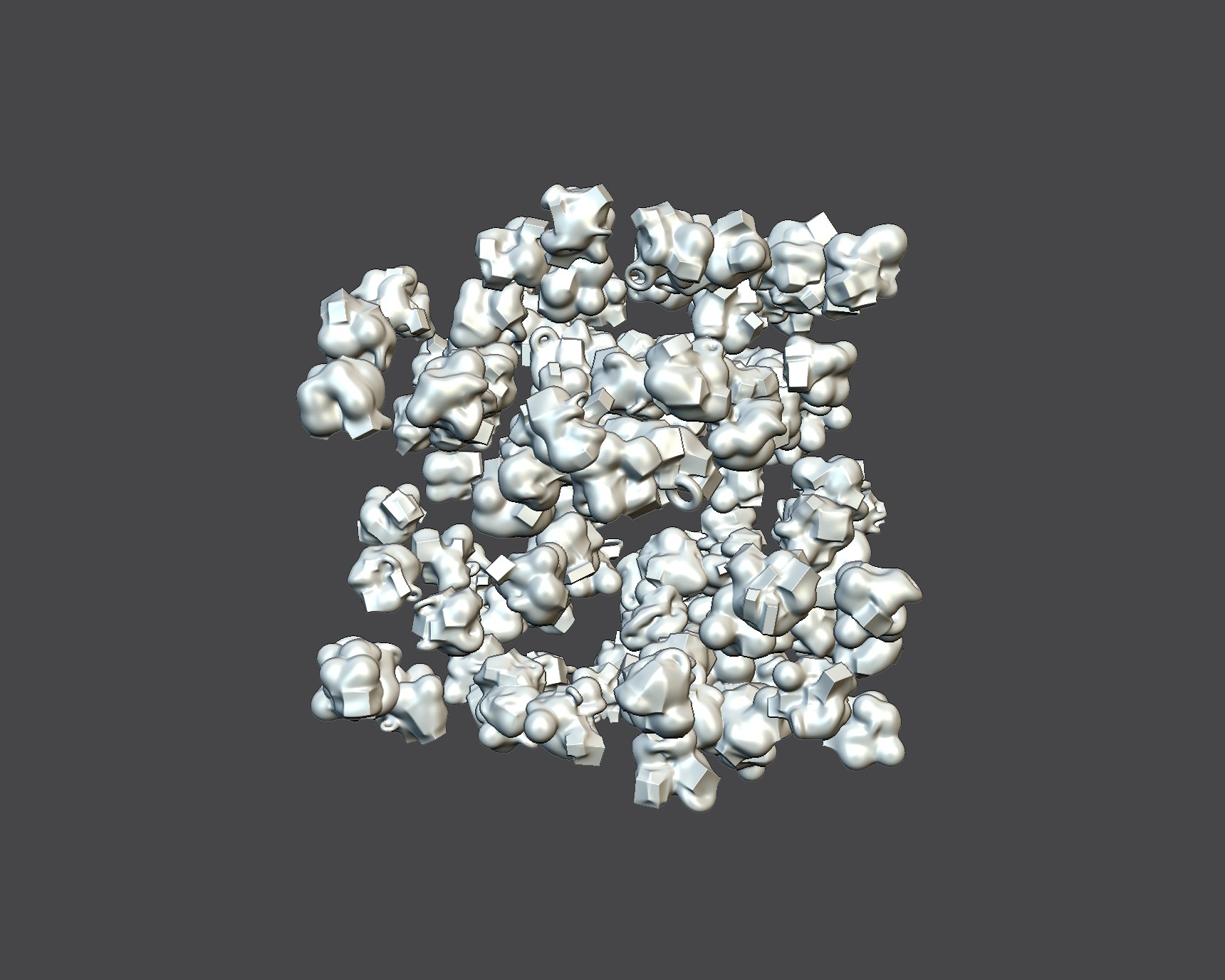}
    
    \includegraphics[width=0.24\linewidth]{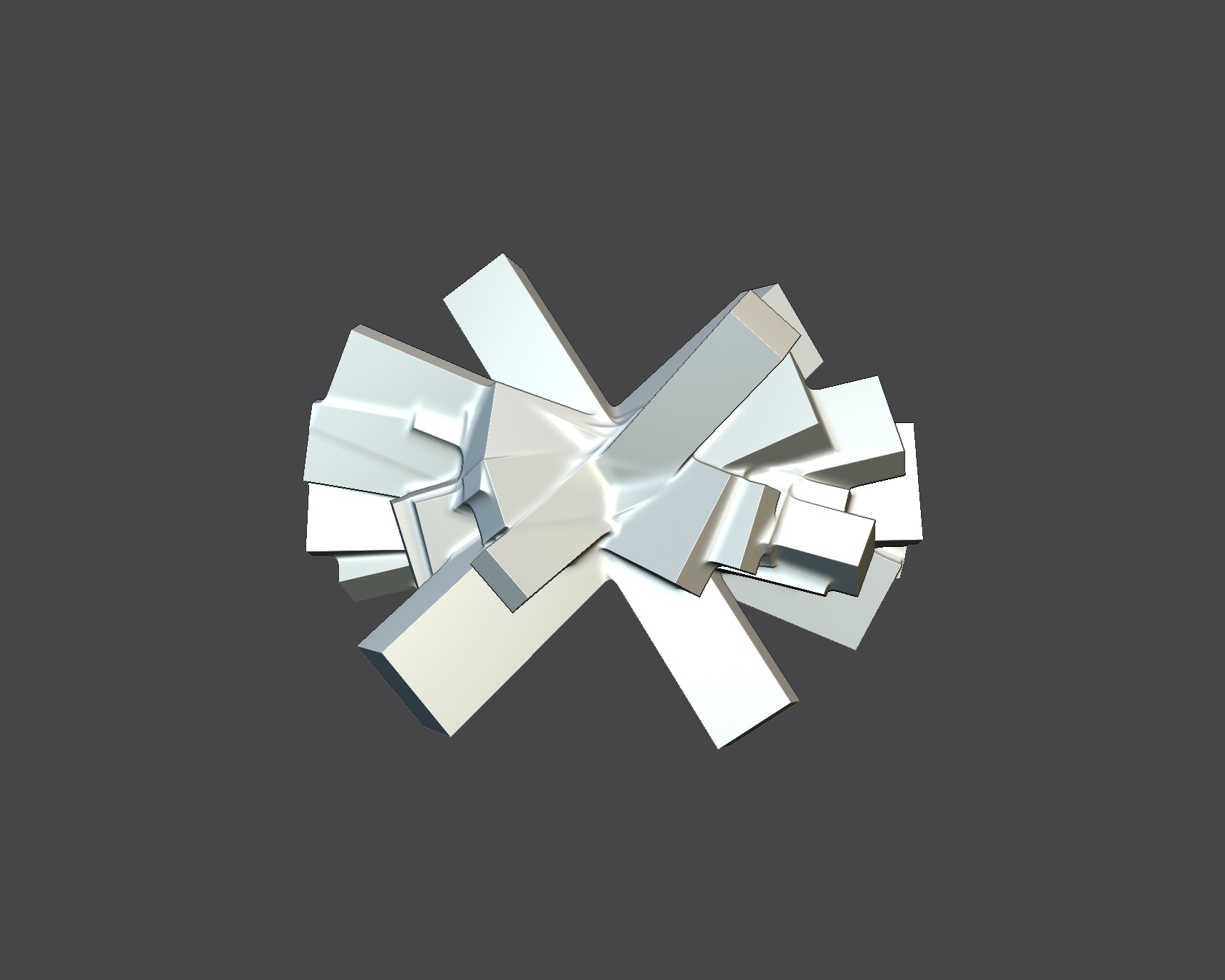}
    \includegraphics[width=0.24\linewidth]{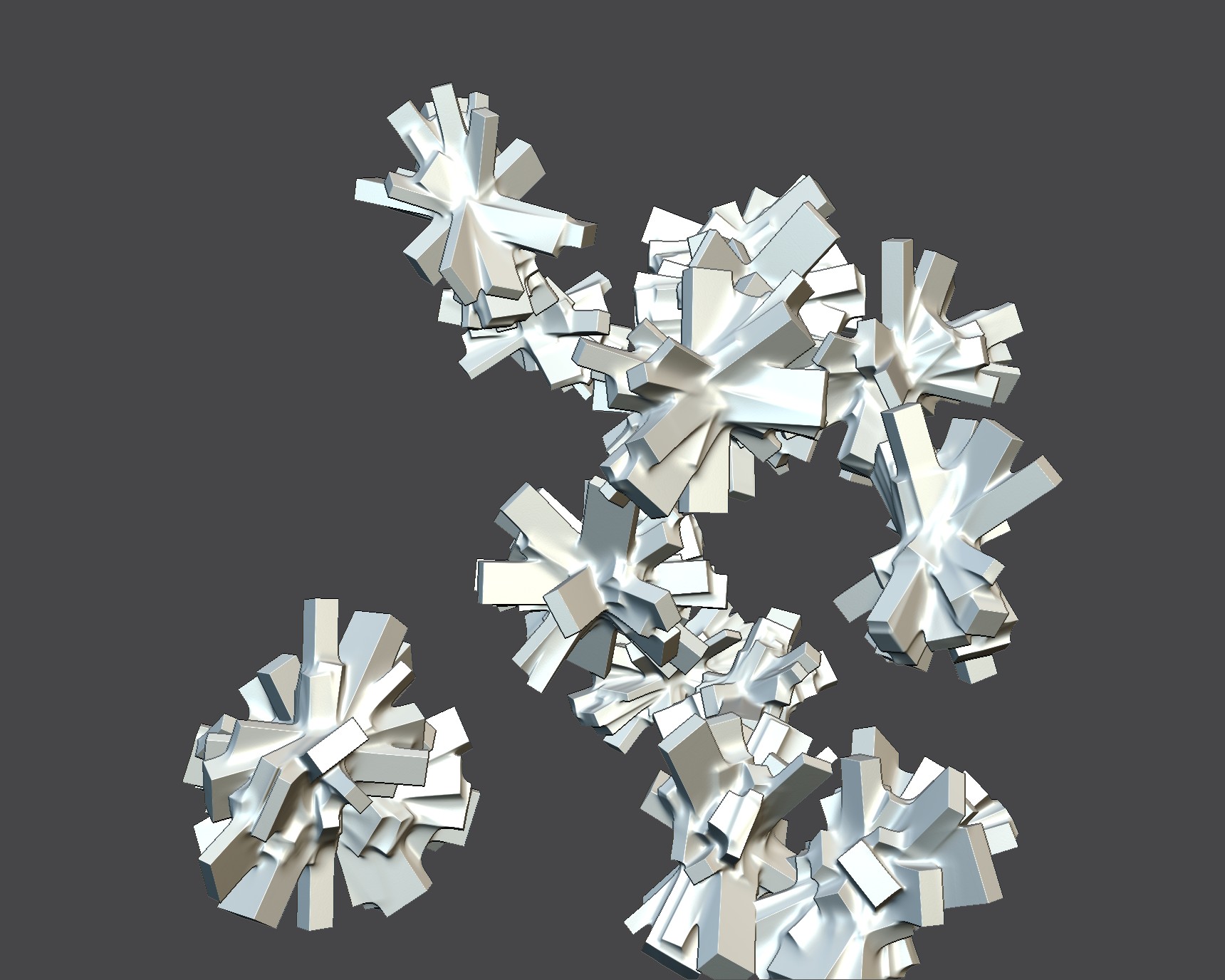}
    \includegraphics[width=0.24\linewidth]{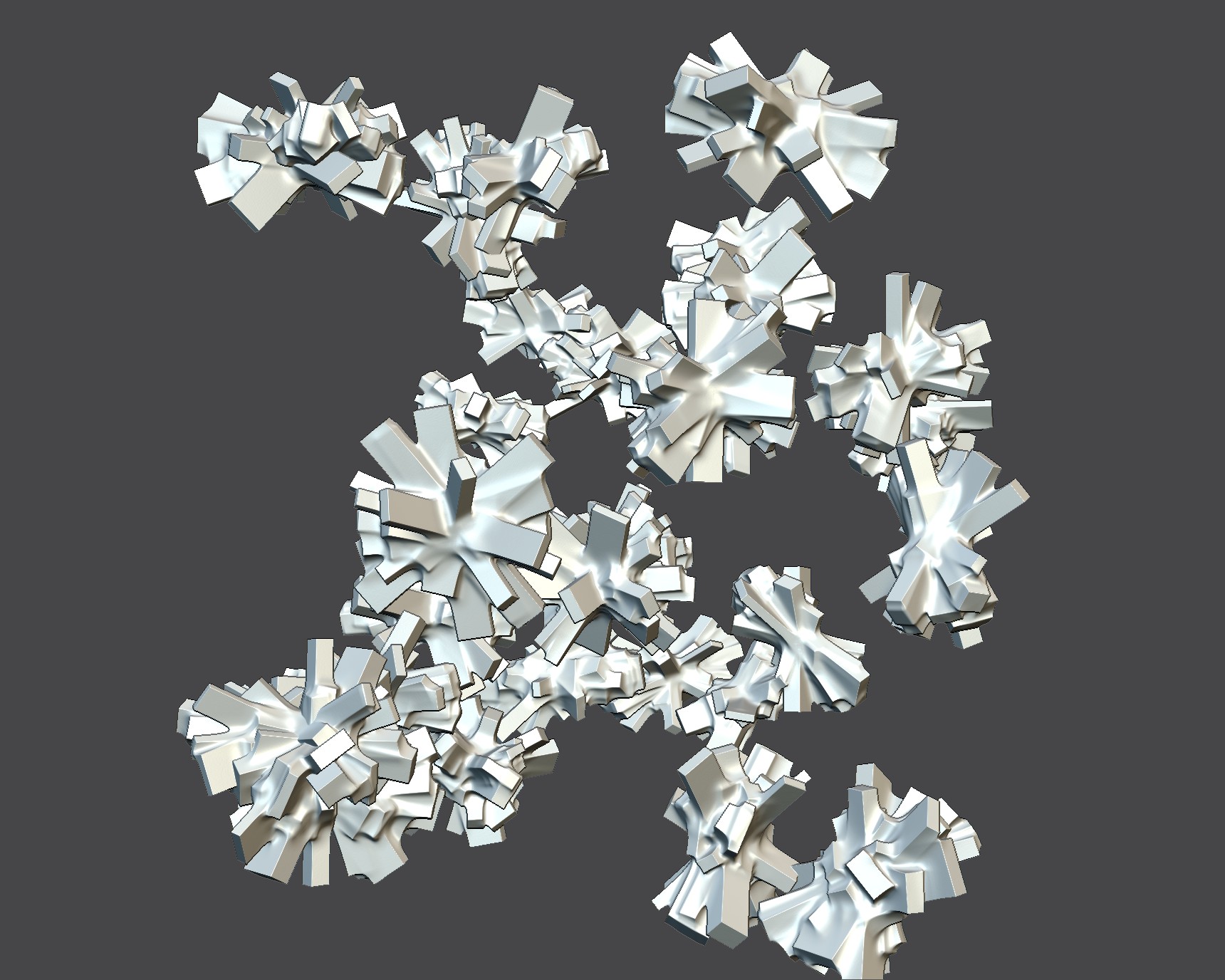}
    \includegraphics[width=0.24\linewidth]{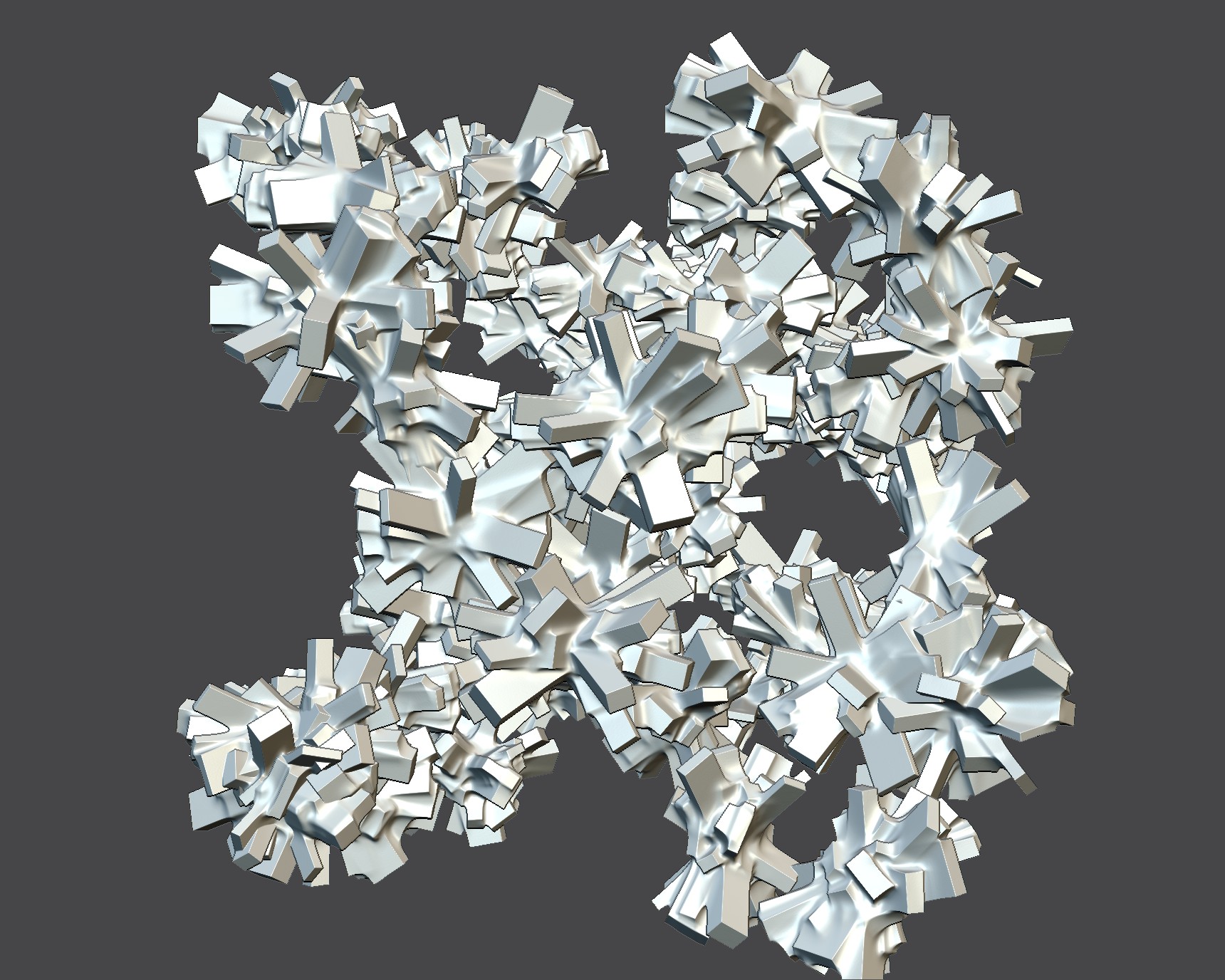}
    
    \includegraphics[width=0.24\linewidth]{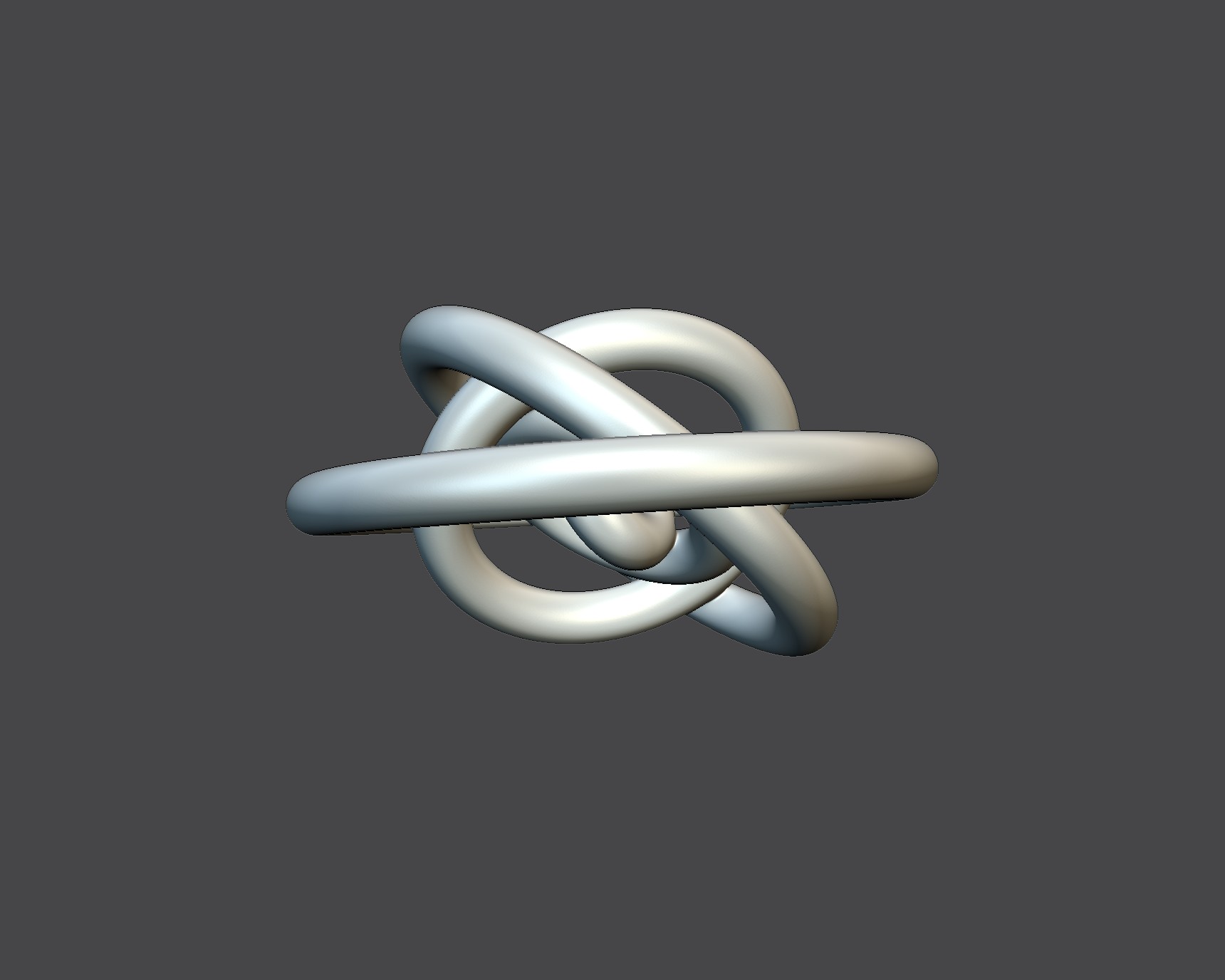}
    \includegraphics[width=0.24\linewidth]{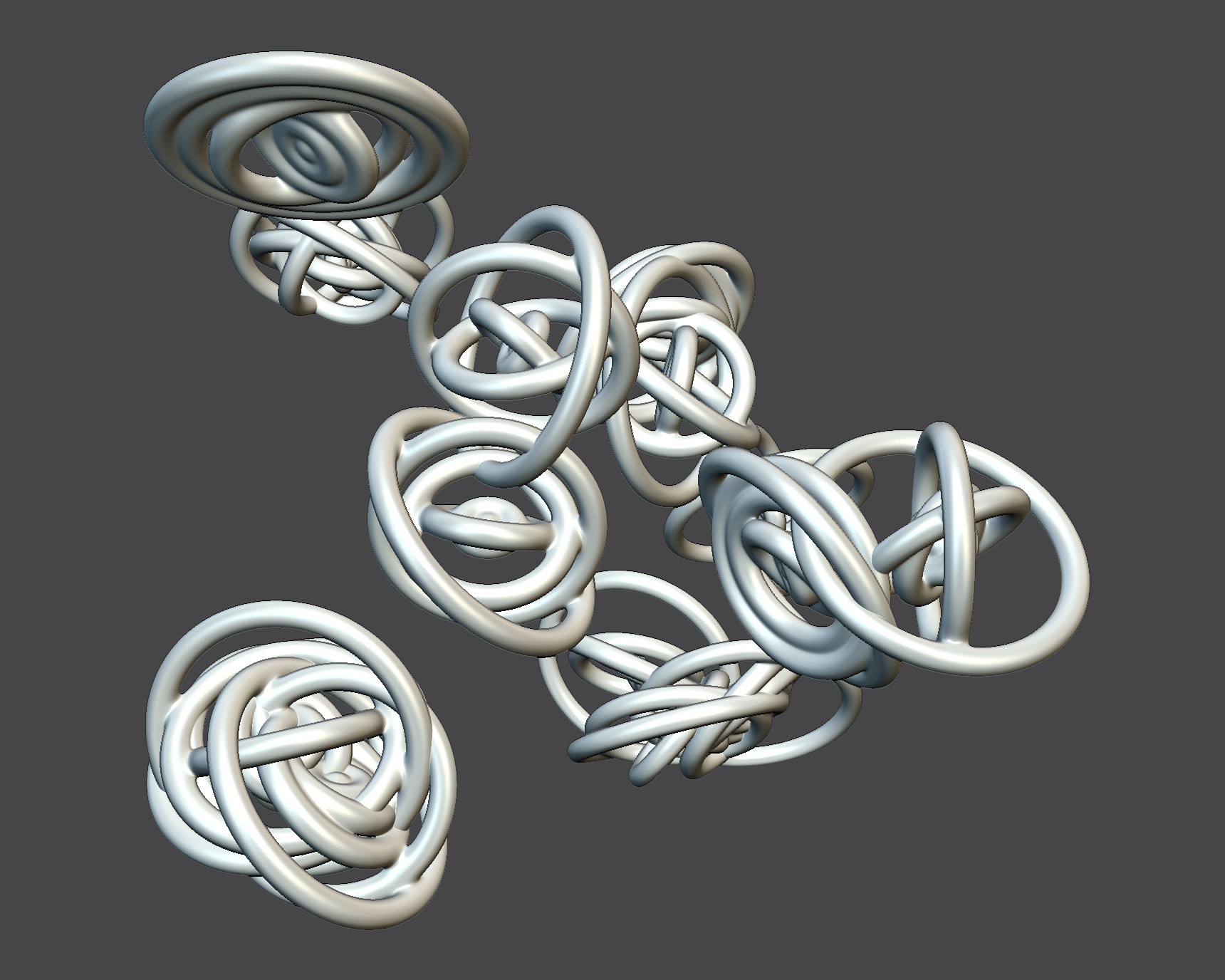}
    \includegraphics[width=0.24\linewidth]{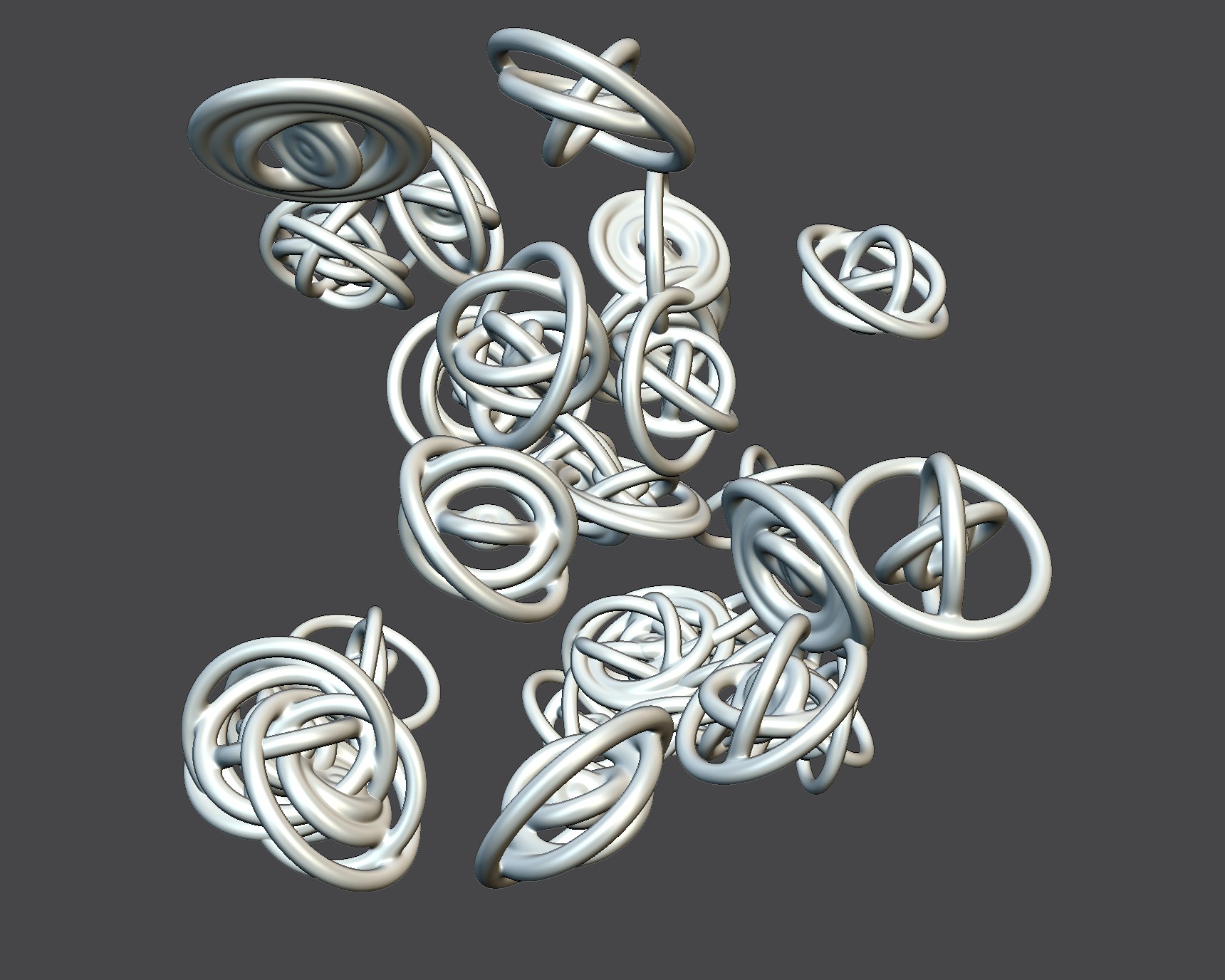}
    \includegraphics[width=0.24\linewidth]{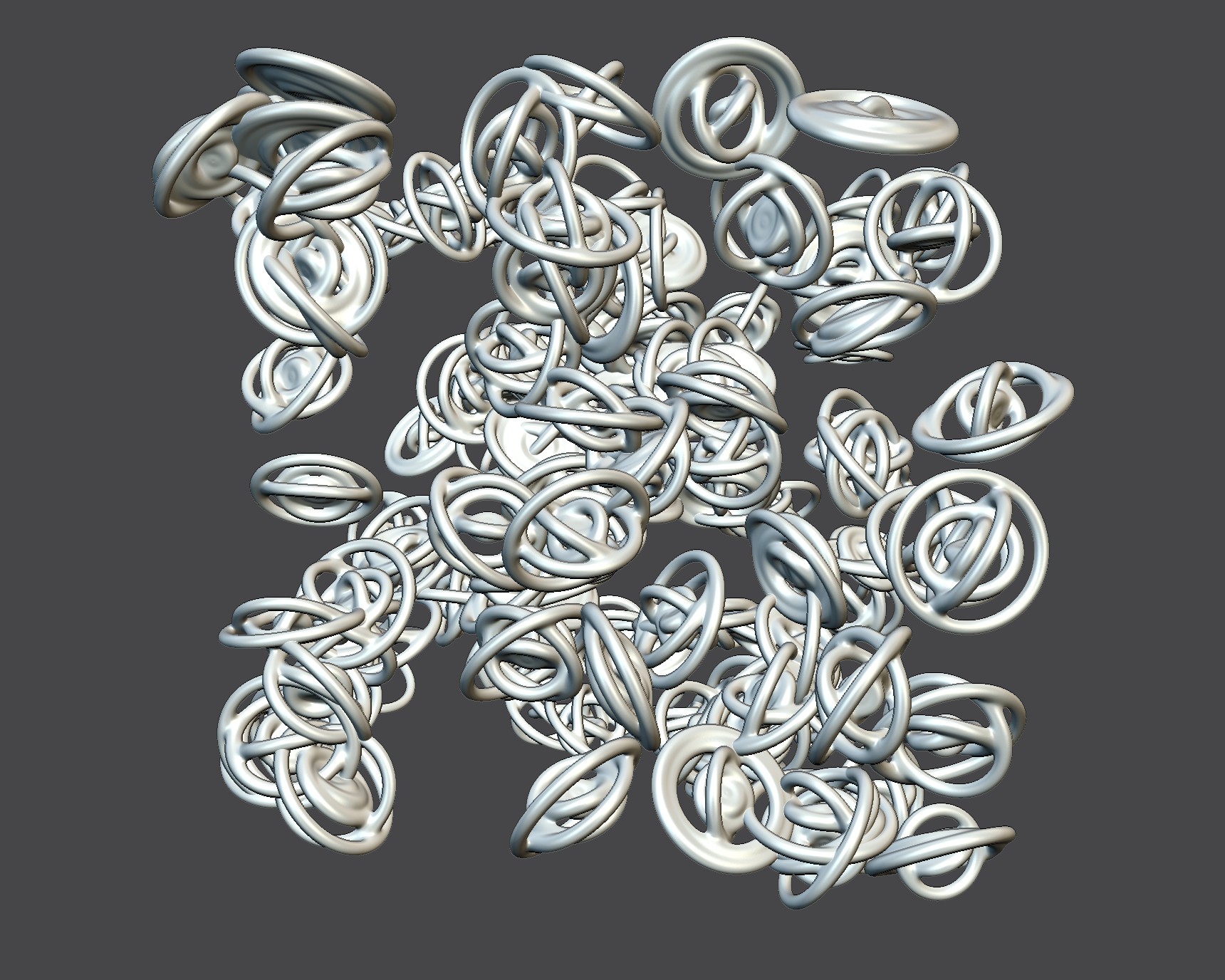}
    
    \includegraphics[width=0.24\linewidth]{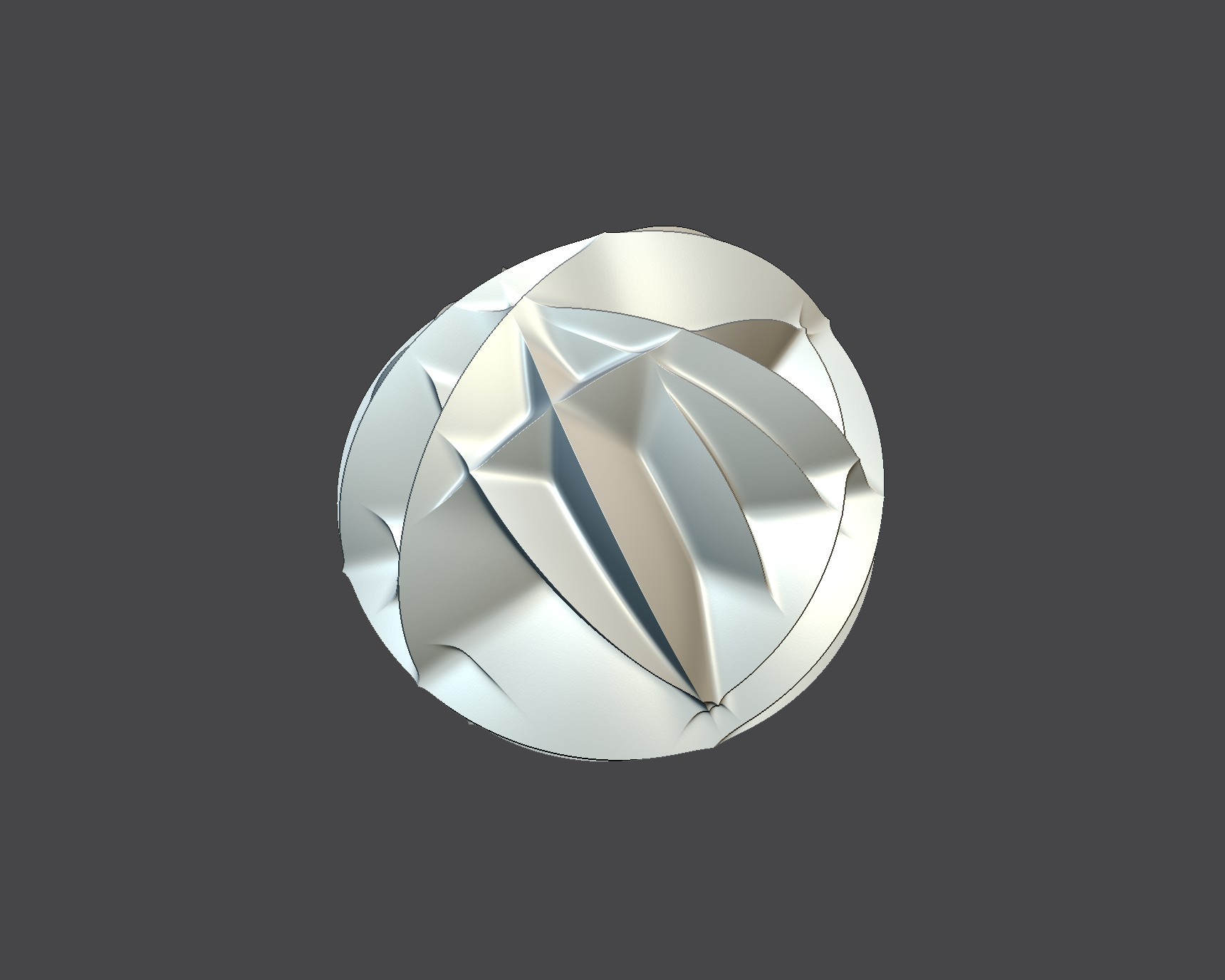}
    \includegraphics[width=0.24\linewidth]{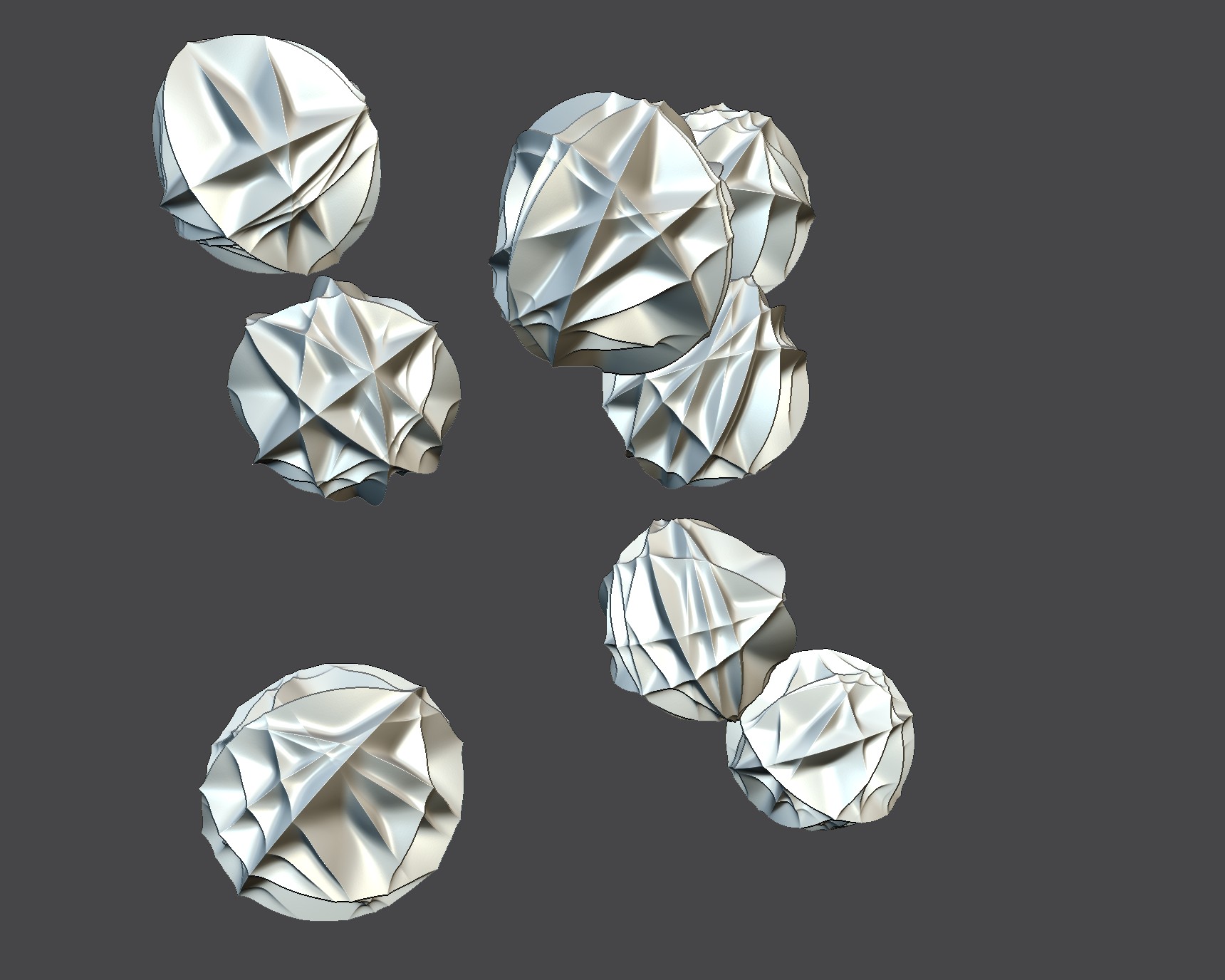}
    \includegraphics[width=0.24\linewidth]{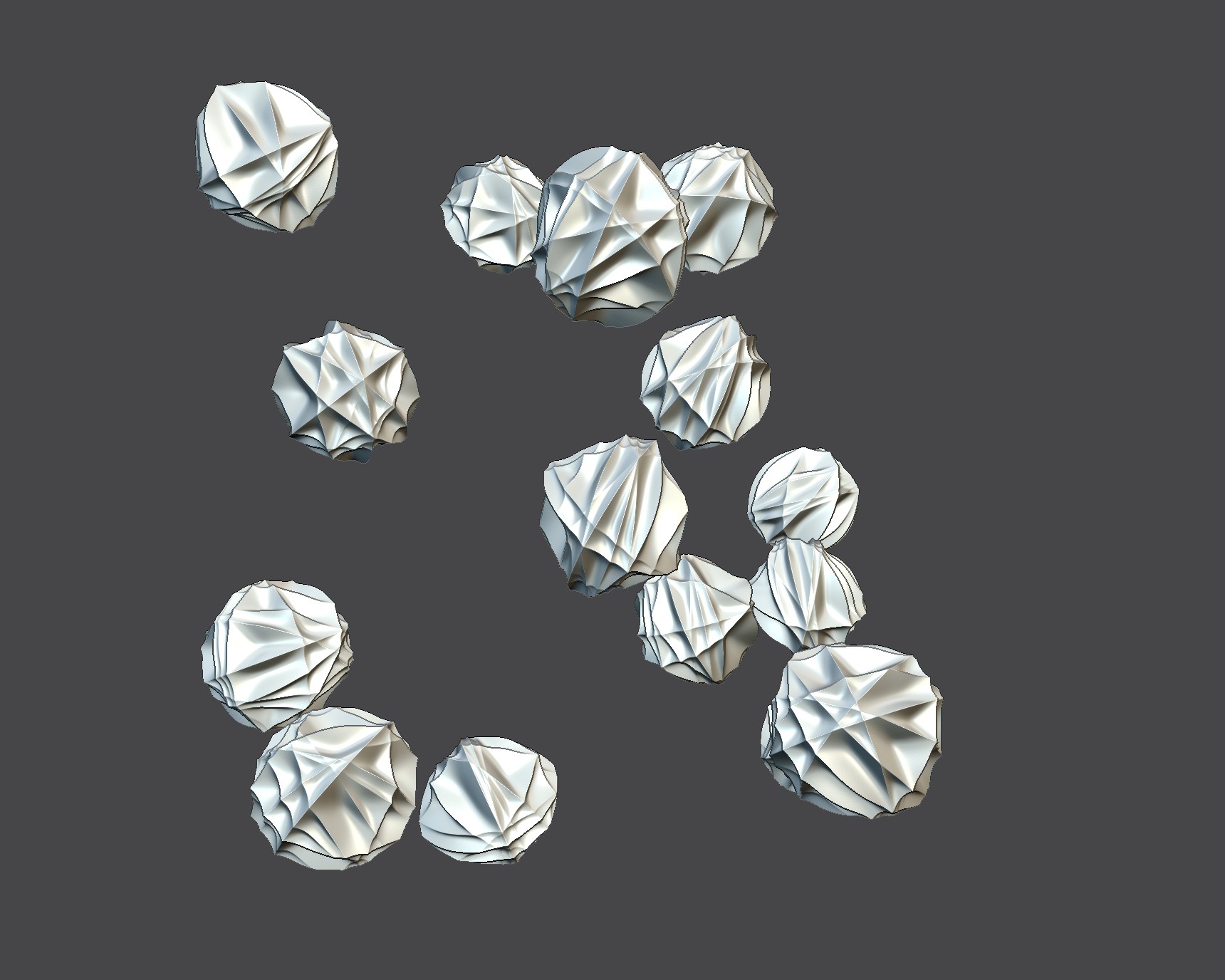}
    \includegraphics[width=0.24\linewidth]{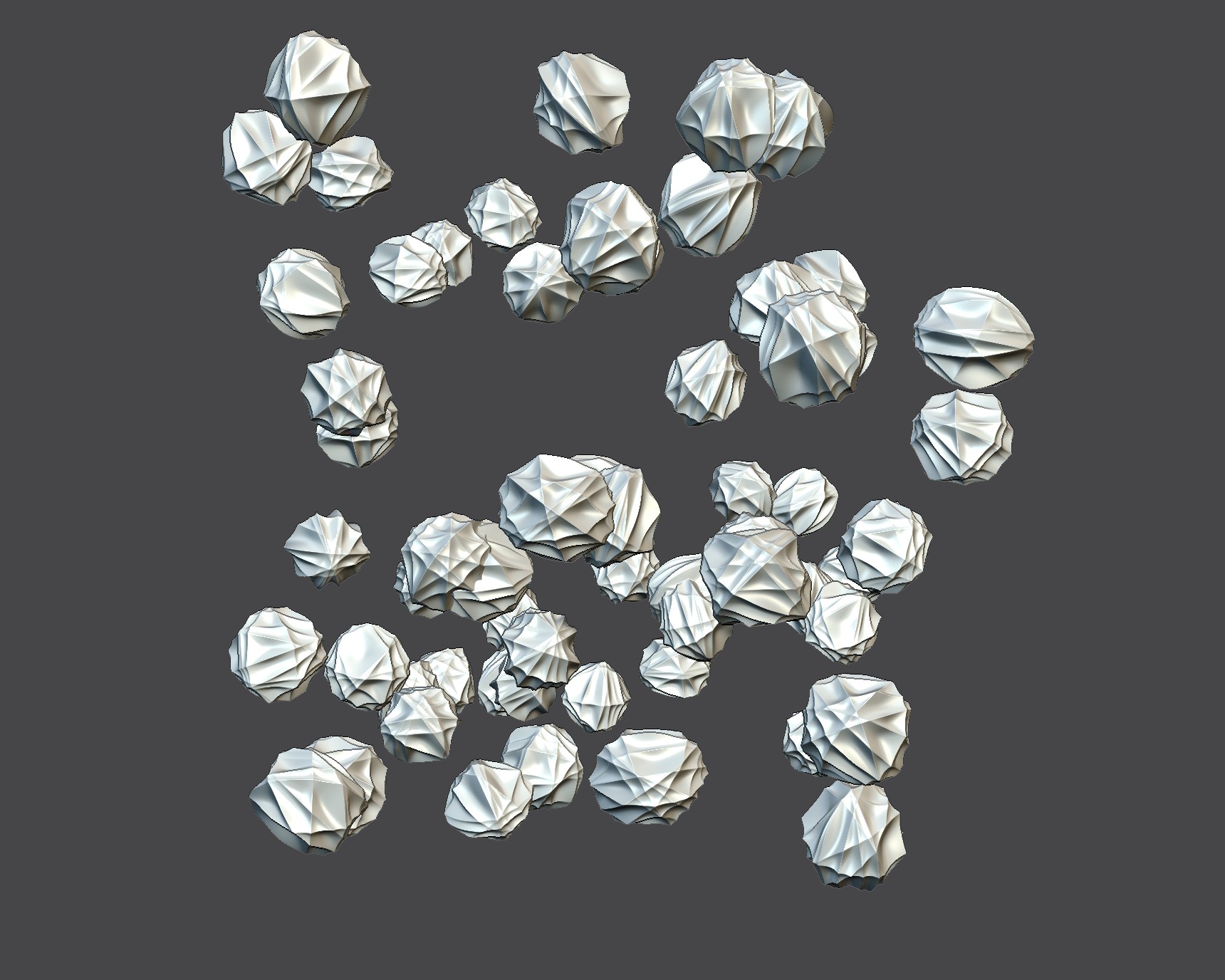}

    \includegraphics[width=0.24\linewidth]{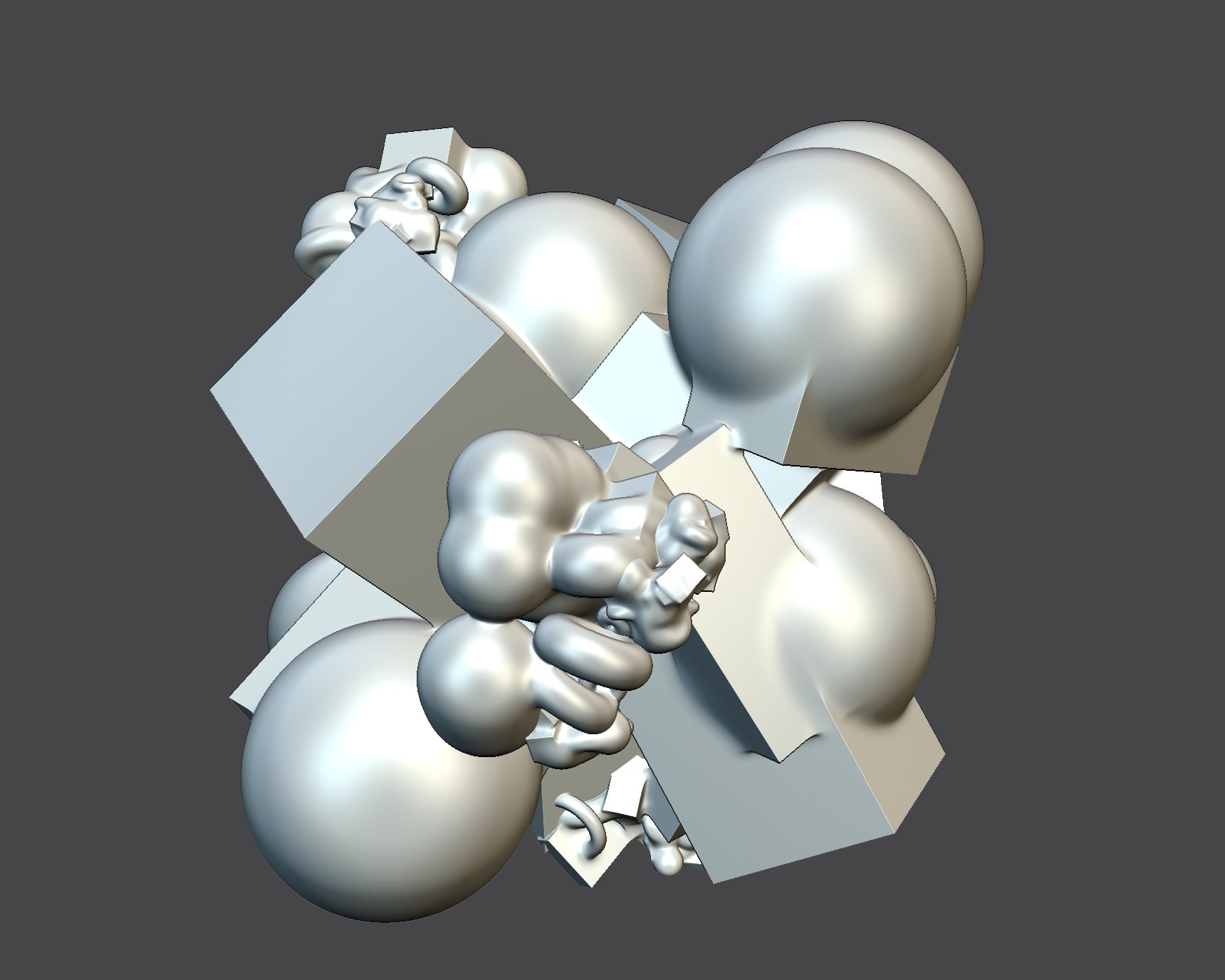}
    \includegraphics[width=0.24\linewidth]{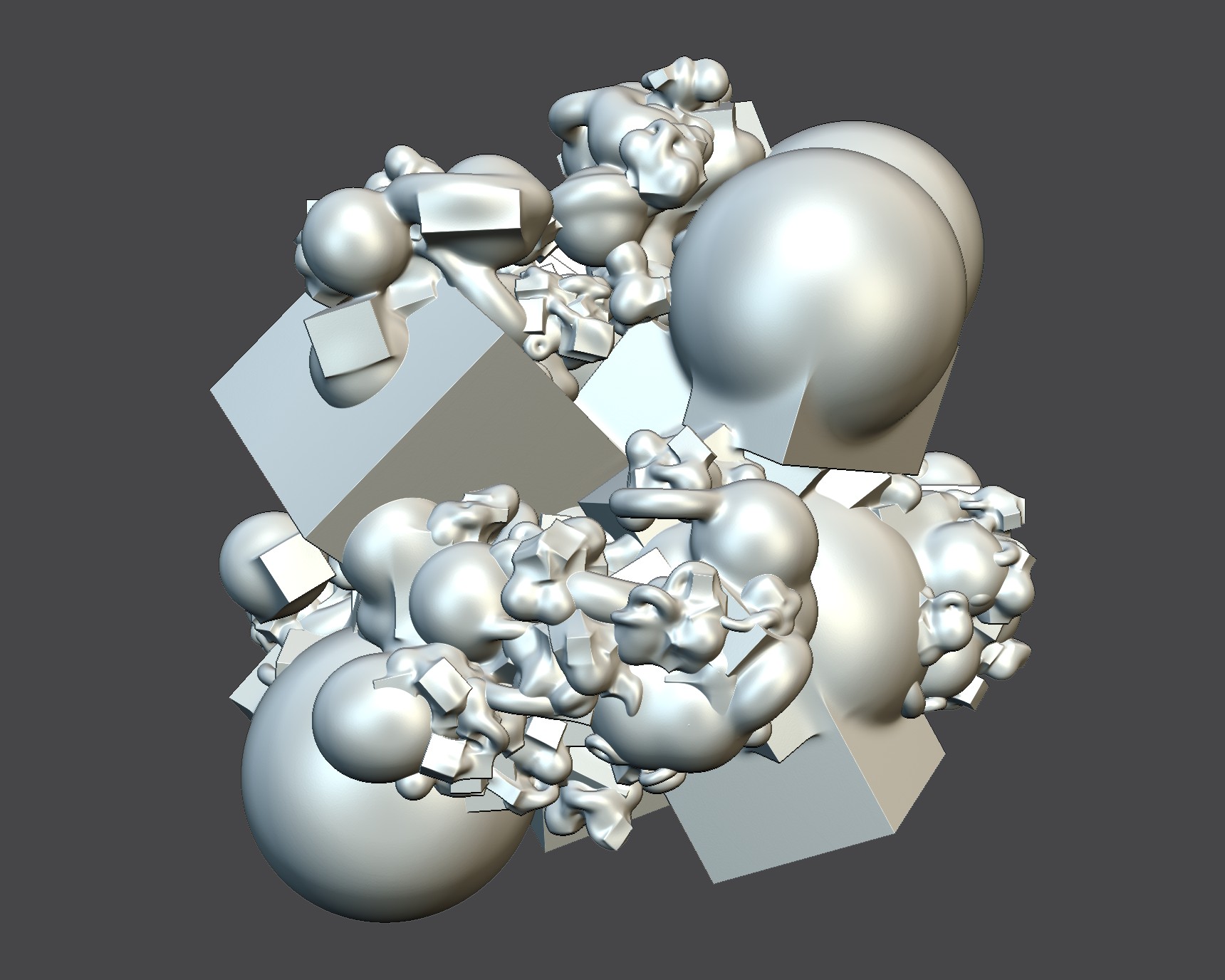}
    \includegraphics[width=0.24\linewidth]{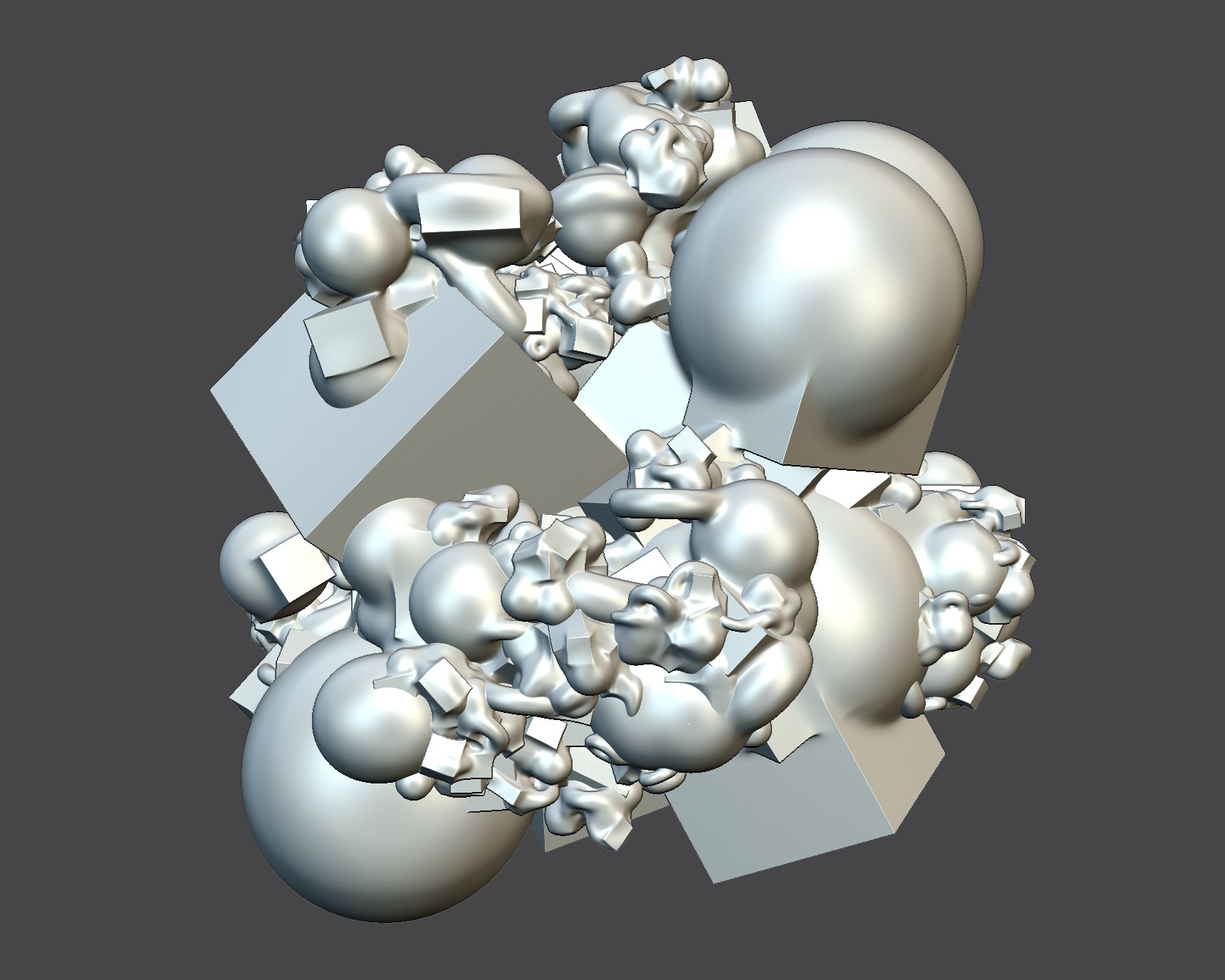}
    \includegraphics[width=0.24\linewidth]{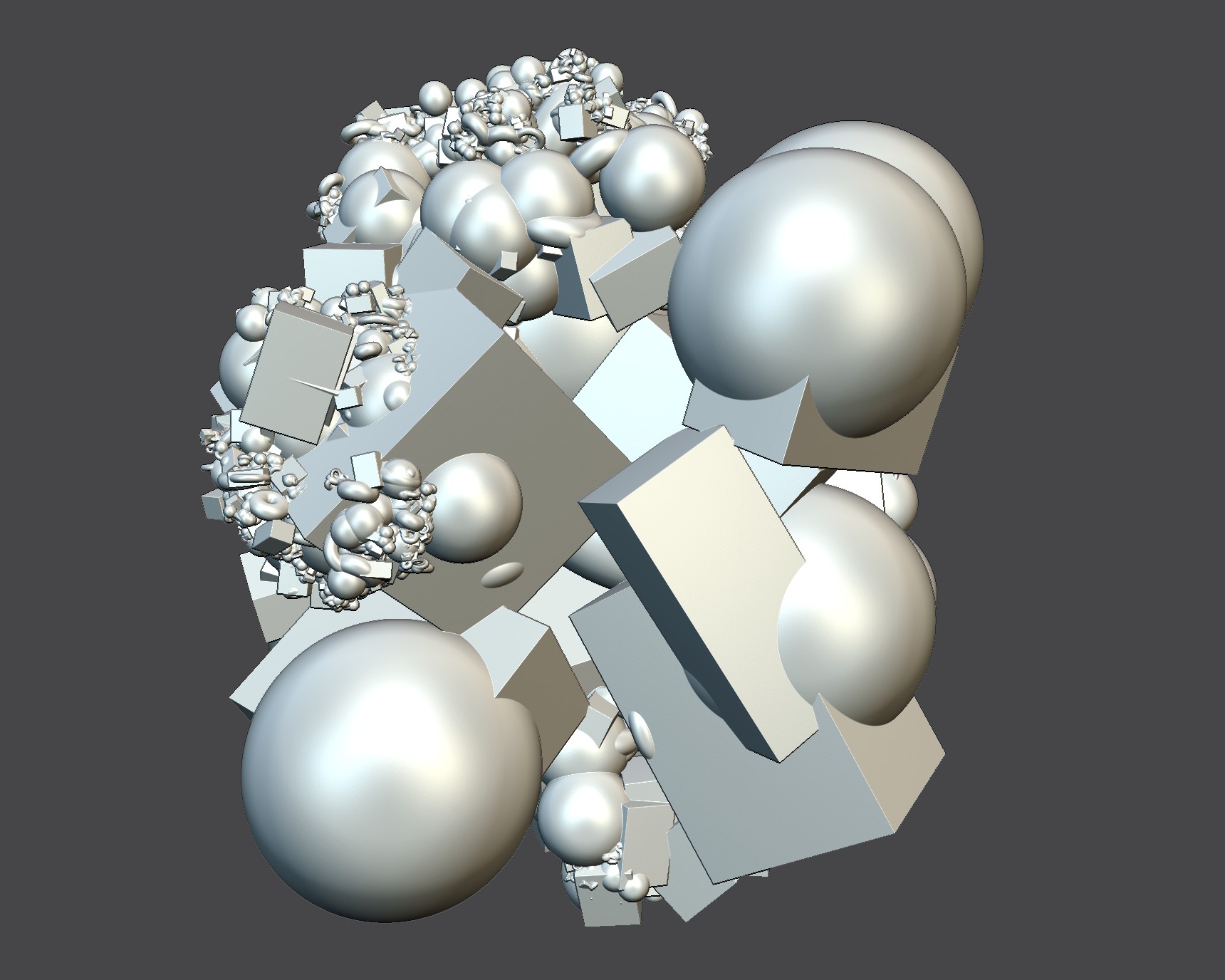}

    \caption{\label{fig:synth-comp} Synthetic examples used for comparison with~\cite{Keeter2020} displayed here with smooth blending (common blend parameter for all scales) with the exception of R4C3-4.
    }
  \end{figure}

%% file: dexel_graph.tex
  \begin{figure}[h]
    \centering
    \includegraphics[width=0.45\linewidth]{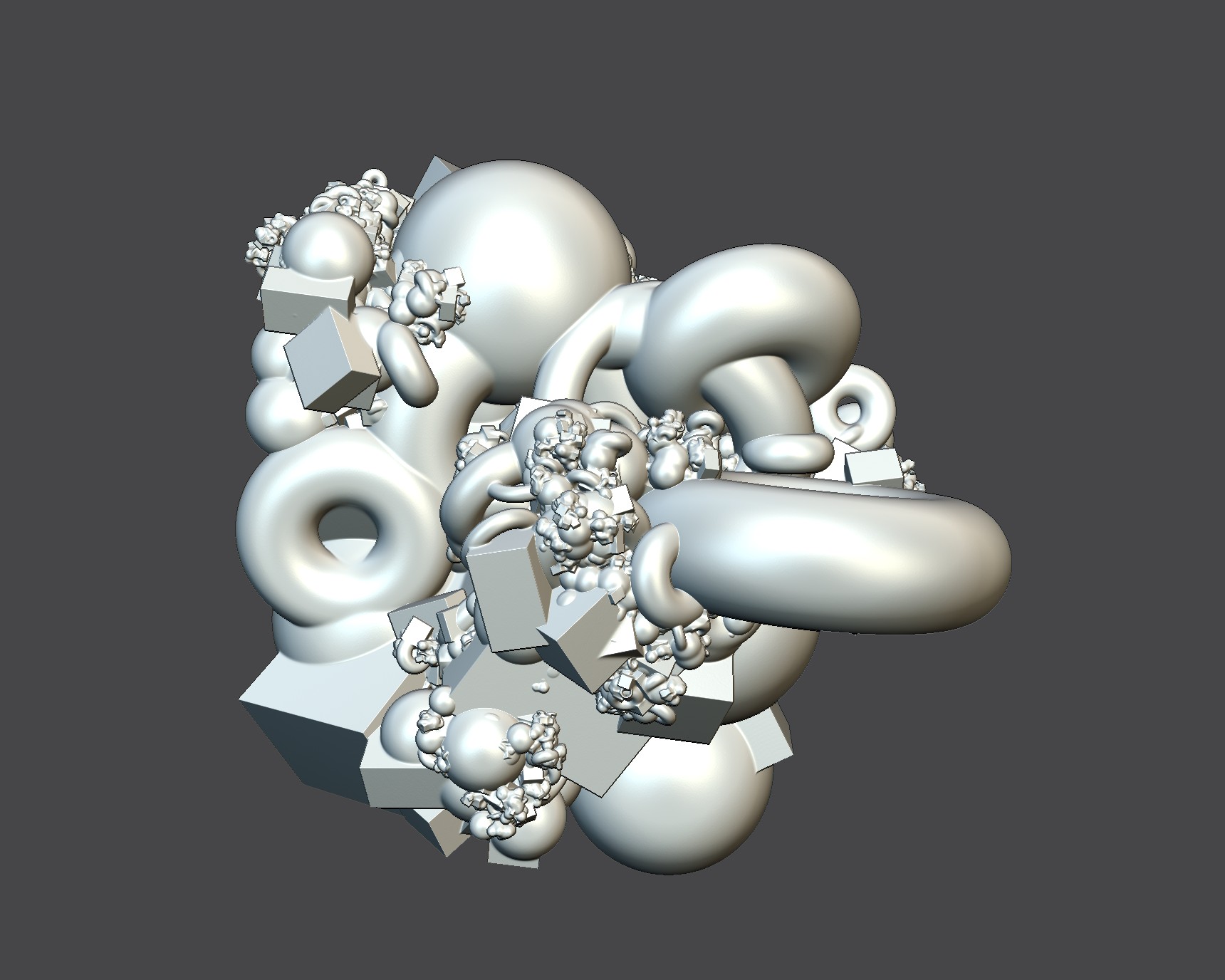}
    \includegraphics[width=0.45\linewidth]{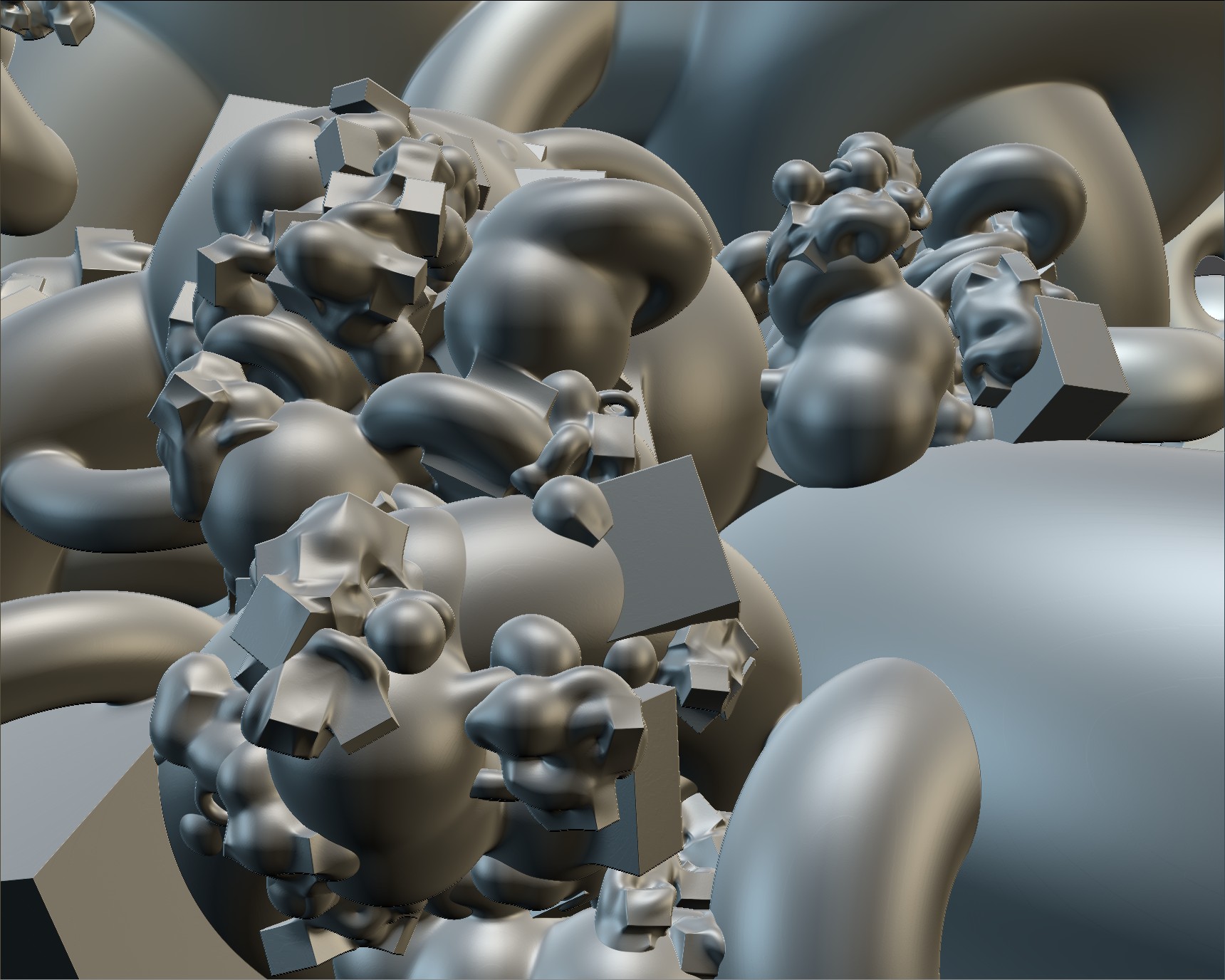}

    \caption{\label{fig:zoom_multi_res} Large number of small primitives with large blend range increase significantly the building time of the A-Buffer. Reducing blend range allow to render properly example of Figure~\ref{fig:synth-comp} R4C3 (22ms per frame for both full view and close up).
    }
  \end{figure}

%% file: appendix_comparison.tex
\input{runtime_table-comparison.tex}

%% file: runtime_table-comparison.tex
\begin{table*}[tbhp]
	\begin{center}
		\begin{tabular}{|l||c||c|c|c|c||c|c|  }
			\hline
			&  &  \multicolumn{4}{|c||}{nVidia Quadro P1000} & \multicolumn{2}{|c|}{nVidia GeForce 2080 RTX} \\
			\cline{3-8}
			& &  \multicolumn{2}{|c|}{min} & \multicolumn{2}{|c|}{smooth} & min & smooth  \\
			\cline{3-8}
			Object &  \# of primitives 
			& Ours & \cite{Keeter2020} & Ours & \cite{Keeter2020} 
			& Ours & Ours  \\
			\hline
\hline
			Fig.\ref{fig:synth-comp}(R0 C0)
			 & 6 
			 & 6.7 &  18.3 & 10.7 & 33.1
			 & 4.1 & 5.2  \\
			 Fig.\ref{fig:synth-comp}(R0 C1)
			 & 96 
			 & 26.3 &  117.7  & 55.3 &  291.9
			 & 5.6 & 6.7 \\
			 Fig.\ref{fig:synth-comp}(R0 C2)
			 & 192 
			 & 27.4 & 155.1 & 67.8 &  482.2
			 & 5.6 & 8.1  \\
			 Fig.\ref{fig:synth-comp}(R0 C3)
			 & 768 
			 & 42.1 & 598.1 & 135.9 & -
			 & 6.8  & 15.7 \\
\hline
			 Fig.\ref{fig:synth-comp}(R1 C0)
			 & 25 
			 & 12.6 & 39.3 &  21.4 &  -
			 & 4.4 & 5.6   \\
			 Fig.\ref{fig:synth-comp}(R1 C1)
			 & 400 
			 & 17.6 & 121 & 43.7 &  -
			 & 6 & 10.2  \\
			 Fig.\ref{fig:synth-comp}(R1 C2)
			 & 800 
			 & 17.7 & 209.4 & 64.5 &  -
			 & 5 & 10.6  \\
			 Fig.\ref{fig:synth-comp}(R1 C3)
			 & 3200 
			 & 22.2 & 584.7 & 188.8 &  -
			 & 5.5 & 20.8   \\
\hline 
			 Fig.\ref{fig:synth-comp}(R2 C0)
			 & 25 
			 & 6.8 & 21.3 & 26.2 & -
			 & 5.2 & 6.6  \\
			 Fig.\ref{fig:synth-comp}(R2 C1)
			 & 400 
			 & 18.6 & 208.5 & 76.4 &  -
			 & 5.8 & 9.4 \\
			 Fig.\ref{fig:synth-comp}(R2 C2)
			 & 800 
			 & 20.8 & 669.4 & 119.7 &  -
			 & 5.6 & 15.2  \\
			 Fig.\ref{fig:synth-comp}(R2 C3)
			 & 1600 
			 & 31.3 & - & 183.8 &  -
			 & 6.6 & 21.3 \\
\hline
			 Fig.\ref{fig:synth-comp}(R3 C0)
			 & 24 
			 & 51.4 & 20.6 & 62.2 &  76.5
			 & 8.7 & 8.9  \\
			 Fig.\ref{fig:synth-comp}(R3 C1)
			 & 192 
			 &  100.4 & 48.2 & 154.9 &   262.9
			 & 12.2 & 18  \\
			 Fig.\ref{fig:synth-comp}(R3 C2)
			 & 384
			 & 90.4 & 57.7 & 167.4 &  401.8
			 & 9.1 & 15.6  \\
			 Fig.\ref{fig:synth-comp}(R3 C3)
			 & 1536 
			 & 134.6 & 140.7 & 320.2 &   -
			 & 14.9  & 32.4   \\
\hline
			 Fig.\ref{fig:synth-comp}(R4 C0)
			 & 1075 
			 & 19.7 & 185.3 & 65.2 & -
			 & 5.5 & 8  \\
			 Fig.\ref{fig:synth-comp}(R4 C1)
			 & 3925 
			 & 26.7 & 650.8 & 118.8 & -
			 & 5.7 & 19.2  \\
			 Fig.\ref{fig:synth-comp}(R4 C2)
			 & 47125 
			 & 42.2 & - & - & -
			 & 6.3 & -  \\
			 Fig.\ref{fig:synth-comp}(R4 C3)
			 & 565525 
			 & 252.8 & - & - & -
			 & 54.6 & -  \\

			\hline
		\end{tabular}
				\caption{Comparison of rendering times for examples of Figure~\ref{fig:synth-comp} (in milliseconds) averaged over 20 frames, at $1024\times 1024$ for the Quadro P1000 and  $1720 \times 1376$ for the RTX 2080. 
				Note that if combined preprocessing and rendering took over 4 minutes, runtime are not provided.}
		\label{tab:runtime_comparison_keeter}
	\end{center}
\end{table*}

\begin{table*}[tbhp]
	\begin{center}
		\begin{tabular}{|l||c|c||c|c||c|c||c|c||  }
			\hline
			&  \multicolumn{8}{|c||}{nVidia Quadro P1000}  \\
			\cline{2-9}
			&  \multicolumn{4}{|c||}{min} & \multicolumn{4}{|c||}{smooth}  \\
			\cline{2-9}
			& \multicolumn{2}{|c|}{Expression data size (kB)} & \multicolumn{2}{|c|}{preprocessing time} 
			& \multicolumn{2}{|c|}{Expression data size (kB)} & \multicolumn{2}{|c|}{preprocessing time}  \\
			\cline{2-9}
			Object 
			& Ours & \cite{Keeter2020} & Ours & \cite{Keeter2020} & Ours & \cite{Keeter2020} & Ours & \cite{Keeter2020}  \\
			\hline
\hline
			Fig.\ref{fig:synth-comp}(R0 C0)
			& 0.4 &  1.2 & 1 & $<1$
			& 0.5 &  1.5 & 1 & 11  \\
			 Fig.\ref{fig:synth-comp}(R0 C1)
			 & 7.2 &  20.3 & 1 & 12
			 & 8.4 &  25.4 & 1 & 232  \\
			 Fig.\ref{fig:synth-comp}(R0 C2)
			 & 14.3 &  40.6 & 2 & 27
			 & 16.9 &  50.9 & 2 & 573  \\
			 Fig.\ref{fig:synth-comp}(R0 C3)
			 & 57.3 &  162.8 & 5 & 132
			 & 67.6 &  203.7 & 5 & 2861  \\
\hline
			 Fig.\ref{fig:synth-comp}(R1 C0)
			 & 1.7 &  5.5 & 2 & 3  
			 & 2.1 &  - & 1 & - \\
			 Fig.\ref{fig:synth-comp}(R1 C1)
			 & 28.0 &  87.36 & 3 & 77 
			 & 34.2 &  - & 4 & - \\
			 Fig.\ref{fig:synth-comp}(R1 C2)
			 & 55.1 &  169.1 & 6 & 160 
			 & 67.4 &  - & 7 & - \\
			 Fig.\ref{fig:synth-comp}(R1 C3)
			 & 220.2 &  - & 25 & -
			 & 269.4 &  - & 21 & -  \\
\hline 
			 Fig.\ref{fig:synth-comp}(R2 C0)
			 & 2.0 &  7.1 & 2 & 4
			 & 2.4 &  - & 1 & -  \\
			 Fig.\ref{fig:synth-comp}(R2 C1)
			 & 32.0 &  122.3 & 5 & 113
			 & 38.1 &  - & 3 & -  \\
			 Fig.\ref{fig:synth-comp}(R2 C2)
			 & 64.0 &  244.5 & 5 & 260
			 & 76.2 &  - & 6 & -  \\
			 Fig.\ref{fig:synth-comp}(R2 C3)
			 & 128.0 &  - & 11 & -
			 & 152.6 &  - & 11 & -  \\
\hline
			 Fig.\ref{fig:synth-comp}(R3 C0)
			 & 1.1 &  - & 1 & 1
			 & 1.3 &  3.1 & 2 & 8066  \\
			 Fig.\ref{fig:synth-comp}(R3 C1)
			 & 9.2 &  19.0 & 3 & 11
			 & 10.6 &  24.6 & 3 & 68119  \\
			 Fig.\ref{fig:synth-comp}(R3 C2)
			 & 18.4 &  38.1 & 3 & 25
			 & 21.2 &  49.3 & 3 & 146480  \\
			 Fig.\ref{fig:synth-comp}(R3 C3)
			 & 73.7 &  153.4 & 10 & 131
			 & 85.0 &  - & 10 & -  \\
\hline
			 Fig.\ref{fig:synth-comp}(R4 C0)
			 & 74.0 &  226.6 & 9 & 230
			 & 90.5 &  - & 8 & -  \\
			 Fig.\ref{fig:synth-comp}(R4 C1)
			 & 271.8 &  - & 29 & -
			 & 332.1 &  - & 27 & -  \\
			 Fig.\ref{fig:synth-comp}(R4 C2)
			 & 3264.0 &  - & 357 & -
			 & 3987.9 &  - & 362 & -  \\
			 Fig.\ref{fig:synth-comp}(R4 C3)
			 & 39210.8 &  - & 4686 & -
			 & 47897.3 &  - & 5231 & -  \\

			\hline
		\end{tabular}
				\caption{Comparison of pre-processing times (in milliseconds) and GPU data size.
				}
		\label{tab:preprocess_comparison_keeter}
	\end{center}
\end{table*}


%% file: conclusion.tex
\section{Conclusion and Future work}

We have introduced a new rendering pipeline for direct visualization of approximate signed distance fields structured in blobtrees. Our pipeline is based on three key ideas, which are 1) compactified blending operators to define volumes of interest, 2) a sparse bottom-up traversal of blobtrees removing the need for preprocessing when only primitive parameters are changing, and 3) coherent processing of rays relying on a low-resolution A-buffer for synchronization.
The combination of those ingredients provides significant runtime improvement compared to state-of-the-art method, especially for large blobtrees.

Our pipeline has two straightforward extensions: usage with density fields and more advanced iterative ray processing, such as segment-tracing. When targeting higher resolution, using larger tiles and processing by depth slabs could also limit memory usage.